\documentclass[aps,prx,reprint,amsmath,amssymb,superscriptaddress,nofootinbib,floatfix]{revtex4-2}

\usepackage[T1]{fontenc}
\usepackage[utf8]{inputenc}
\usepackage{amsthm}
\usepackage{mathtools}
\usepackage{bbm}
\usepackage{physics}
\usepackage{graphicx}
\usepackage{booktabs}
\usepackage{xcolor}
\usepackage{microtype}
\usepackage[hypertexnames=false]{hyperref}
\usepackage[capitalize,nameinlink]{cleveref}
\usepackage{tocloft}

\cftsetpnumwidth{2em}
\cftsetrmarg{2em}
\crefname{section}{Sec.}{Secs.}

\definecolor{red4}{Hsb}{0,0.79,0.72}
\definecolor{blue4}{Hsb}{240,0.65,0.85}
\definecolor{purple4}{Hsb}{330,0.79,0.70}

\allowdisplaybreaks
\hypersetup{
    colorlinks,
    citecolor=blue4,
    linkcolor=red4,
    urlcolor=purple4,
    breaklinks=true,
}

\newcommand{\F}{\mathbb{F}}
\newcommand{\cH}{\mathcal{H}}

\newcommand{\Span}{\mathrm{span}}

\newtheorem{theorem}{Theorem}

\makeatletter
\newcommand{\apptocfile}{atoc}
\let\apptoc@orig@appendix\appendix
\renewcommand{\appendix}{%
  \apptoc@orig@appendix
  \let\apptoc@orig@addtocontents\addtocontents
  \long\def\addtocontents##1##2{%
    \def\apptoc@ext{##1}%
    \def\apptoc@toc{toc}%
    \ifx\apptoc@ext\apptoc@toc
      \apptoc@orig@addtocontents{\apptocfile}{##2}%
    \else
      \apptoc@orig@addtocontents{##1}{##2}%
    \fi
  }%
}
\newcommand{\appendixtableofcontents}{%
  \begingroup
    \hyphenpenalty=5000
    \exhyphenpenalty=5000
    \setcounter{tocdepth}{2}%
    \phantomsection
    \let\addcontentsline\@gobblethree
    \section*{Contents}%
    \pdfbookmark[1]{Appendices}{apxcontents}%
    \let\l@subsubsection\@gobbletwo
    \@starttoc{\apptocfile}%
  \endgroup
}
\let\oldnumberline\numberline
\renewcommand{\numberline}[1]{%
  \if\relax\detokenize{#1}\relax
    \oldnumberline{#1}%
  \else
    \oldnumberline{#1.}%
  \fi
}
\renewcommand\paragraph{\@startsection{paragraph}{4}{\z@}%
  {1ex \@plus1ex \@minus.2ex}%
  {-1em}%
  {\normalfont\normalsize\bfseries}%
}%
\makeatother

\begin{document}

\title{Measurement-Based Uncomputation from an Error Correction Perspective}

\newcommand{\oxddress}{\affiliation{Department of Materials, University of Oxford, Parks Road, Oxford OX1 3PH, United Kingdom}}

\newcommand{\oxmathsaddress}{\affiliation{Mathematical Institute, University of Oxford, Woodstock Road, Oxford OX2 6GG, United Kingdom}}

\newcommand{\oxengaddress}{\affiliation{Department of Engineering Science, University of Oxford, Parks Road, Oxford OX1 3PJ, United Kingdom}}

\newcommand{\impaddress}{\affiliation{Department of Computing, Imperial College London, 180 Queen’s Gate, London SW7 2AZ, United Kingdom}}

\newcommand{\qmaddress}{\affiliation{Quantum Motion, 9 Sterling Way, London N7 9HJ, United Kingdom}}

\let\svthefootnote\thefootnote

\author{Minjun Jeon}
\email{minjun.jeon@materials.ox.ac.uk}
\oxddress
\qmaddress

\author{Po-Wei Huang}
\email{po-wei.huang@maths.ox.ac.uk}
\oxmathsaddress
\qmaddress

\author{Zhenyu Cai}
\email{cai.zhenyu.physics@gmail.com}
\impaddress
\oxengaddress
\qmaddress

\date{September 25, 2026}

\begin{abstract}
    Coherent uncomputation can be as costly as the computation itself, creating substantial spacetime overheads that can be prohibitive on early fault-tolerant quantum computers. We develop an algebraic framework for measurement-based uncomputation (MBU) through a quantum error-correction picture: measurements of the garbage register induce relative phases analogous to errors causing non-trivial syndromes. Accepting only zero-syndrome outcomes achieves post-selected MBU, removing the need for an uncomputation oracle at the cost of a probabilistic success rate determined by the affine rank of the populated garbage support. This rank can be much smaller than the number of garbage qubits, as illustrated through an example of the HHL algorithm. 
    We then derive a structural condition under which the original uncomputation oracle can be reused to correct the phase for all measurement outcomes. No correction is required for the zero syndrome, while corrections for non-zero syndromes can be further adapted to individual measurement outcomes for simpler circuits, as illustrated by existing modular-arithmetic, QROM, and QRAM MBU constructions. Phase-correction circuits can be further simplified by co-optimising their implementation with the representation of the measurement-induced phases. Selecting which syndrome classes to correct interpolates between post-selected, partially corrected, and deterministic MBU, enabling trade-offs among correction cost, workspace, and repetitions.
    
\end{abstract}

\maketitle

\section{Introduction}

Recent experimental progress in quantum error correction, including repeated logical error correction~\cite{Paetznick2024LogicalQubits,Acharya2024BelowSurfaceThreshold},
logical teleportation~\cite{RyanAnderson2024LogicalTeleportation}, and logical gate-set demonstrations~\cite{Zhang2025UniversalLogicalGateSet}, brings the practical realisation of early fault-tolerant quantum computing much closer to reality. This motivates quantum algorithm designs tailored to the resource constraints of early fault-tolerant (EFT) devices. In this regime, the cost of a quantum algorithm depends not only on its gate count, but also on its coherent depth and the number of ancillary qubits that must remain live throughout the computation~\cite{Litinski2019}. Reducing the lifetime and processing cost of temporary quantum workspaces is therefore a critical aspect of practical algorithm design.

Uncomputation is an integral step in quantum computing for releasing these temporary workspaces~\cite{bennett1973logical,Schuchman_2006}. In a compute--use--uncompute pattern, temporary values are generated, used in subsequent operations, and then erased by reversing the computation that produced them. This makes the workspace reusable by disentangling the registers while preserving the useful output, but the inverse circuit can be as expensive as the forward computation. Recent resource improvements to quantum algorithms such as linear-system and differential-equation algorithms~\cite{childs2017quantum,berry2017quantum}, signal-processing and singular-value-transformation methods~\cite{dong2024multilevelquantumsignalprocessing,dong2024feedforwardquantumsingularvalue}, and low-depth estimation schemes~\cite{LiNiYing2023RMPE,huang2026low-depth,erle2026nearly} do not by themselves remove the need to disentangle temporary workspaces in cases where such workspaces are used. For early fault-tolerant implementation, this can become a serious systems-level burden: large ancilla blocks stay live for too long, and the deep inverse circuit often contributes little algorithmic value beyond restoring a clean workspace. This resource requirement for uncomputation has also motivated compiler-level ancilla-reuse and uncomputation tooling, from space-constrained synthesis and high-level framework support~\cite{Seidel_2023,Paradis_2024} to strategic ancilla reuse~\cite{Ding_2020} and automated QROM compilation~\cite{phalak2022optimizationquantumreadonlymemory,sinha2022automatedquantummemorycompilation}.

Measurement-based uncomputation (MBU) offers an often cheaper alternative route.
Instead of coherently reversing the garbage-generating circuits, one measures the garbage qubits and compensates for the induced relative phases on the surviving main register.
This idea already appears in several important settings, including temporary logical-AND constructions~\cite{Gidney_2018}, modular arithmetic circuits~\cite{Luongo2024MBU,luongo2025optimizedcircuitswindowedmodular}, table-lookup methods~\cite{Gidney2019Windowed,luongo2025optimizedcircuitswindowedmodular}, and a QRAM compilation subroutine in neutral atom quantum computers~\cite{Cesa2025FastQRAM}. These constructions motivate the need for a common algebraic framework describing which measurement outcomes preserve the output, which require correction, and how those corrections compare with coherent uncomputation in terms of implementation costs. Such a description should also include protocols that reject expensive-to-correct outcomes rather than requiring every measurement result to be accepted.

\begin{figure*}
    \centering
    \includegraphics[width=0.9\linewidth]{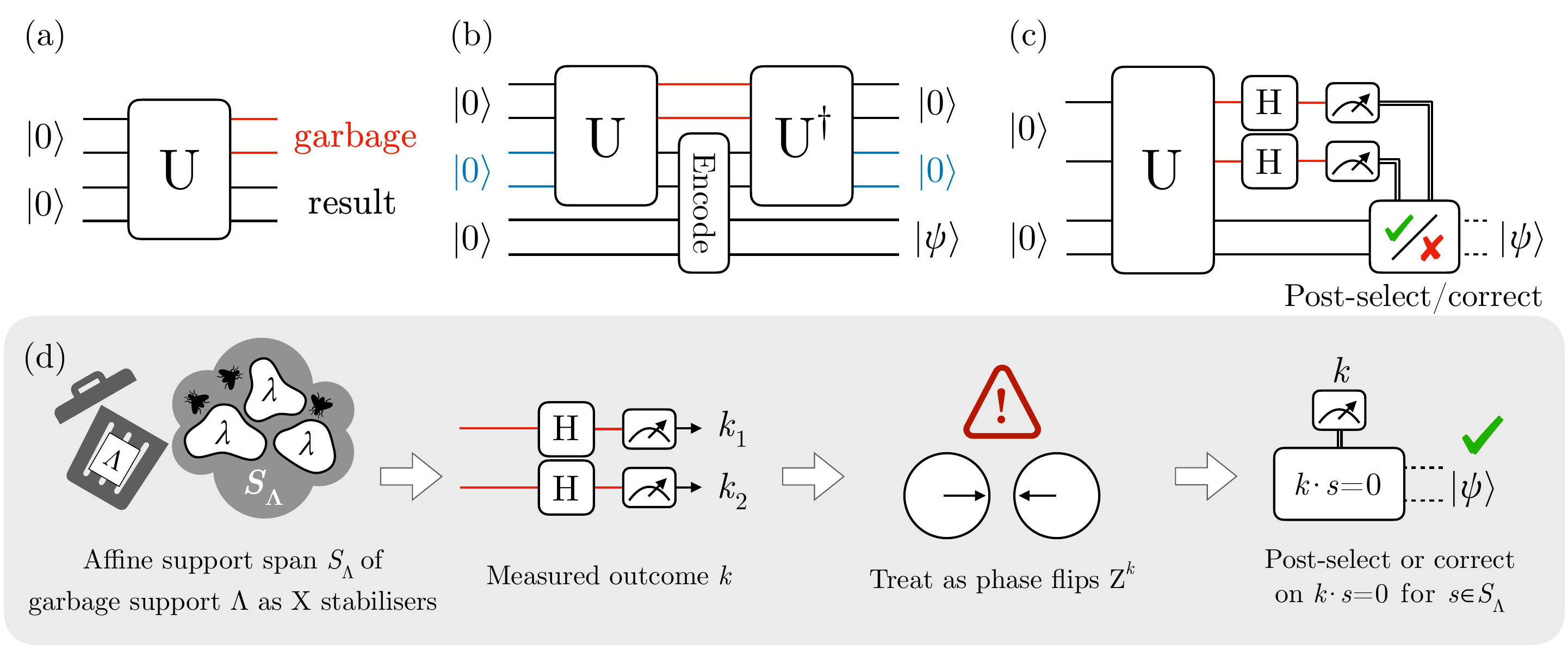}
    \caption{\emph{Conceptual overview of measurement-based uncomputation (MBU) as quantum error correction (QEC).} (a) Standard reversible computation generates a useful result together with a garbage register. (b) Conventional uncomputation restores the garbage register by coherently applying the inverse circuit $U^\dagger$, requiring the garbage qubits to remain coherent until the end of the computation. Before the uncomputation, the result (usually) is copied or encoded into a separate register. (c) In MBU, the garbage register is measured in the $X$ basis instead of being coherently uncomputed. The measurement outcome is accepted only if it satisfies the required acceptance condition (pure post-selection in \cref{sec:framework} or, more generally, the accepted-syndrome criterion of \cref{sec:correctable_framework}), in which case the desired output state is recovered up to a global phase. (d) Quantum error-correction interpretation of MBU. The populated garbage support defines an affine support span $S_\Lambda$, which is interpreted as a space of $X$-stabilisers. Measuring the garbage register produces a phase error $Z^k$ labelled by the measurement outcome $k$. In the pure post-selection regime, successful runs correspond to syndrome-zero outcomes $k\in S_\Lambda^\perp$, while \cref{sec:correctable_framework} generalises this to the acceptance of additional correctable syndrome classes.}
    \label{figMain}
\end{figure*}

Our goal in this paper is to provide that framework. The central idea is to recast MBU as a problem in quantum error correction (QEC)~\cite{Gottesman1997Stabilizer,nielsen2010quantum}. Measuring the garbage register in the $X$ basis is interpreted as introducing a phase error, while the populated garbage support defines an $X$-stabiliser space. Within this picture, the relative phases induced by the measurement outcomes are precisely the corresponding error syndromes, and successful uncomputation either accepts a zero-syndrome outcome or applies a phase error correction associated with a nonzero syndrome.

This QEC perspective leads to an algebraic description of MBU post-selected on the zero syndrome in \cref{thm:succ-prob-post-selected-mbu}. There we show that the post-selection success probability is determined entirely by the affine rank of the populated garbage states, rather than decaying exponentially on the total number of garbage qubits. When this rank is small, replacing a deep coherent inverse circuit with measurement and post-selection can offer a favourable trade-off between coherent depth and repetitions even for a large problem size. We apply this framework to the Harrow--Hassidim--Lloyd (HHL) algorithm~\cite{Harrow_2009}, replacing the inverse quantum phase estimation block with purely post-selected MBU and identifying structured spectral regimes in which the sampling overhead is set by the input bandwidth rather than the system size. 

Building upon this pure post-selection framework, we then ask when MBU can be made fully deterministic. \cref{thm:branch-controlled-deterministic} gives a sufficient condition for using the original uncomputation oracle to correct for all non-zero syndromes. Under this condition, the oracle therefore provides a guaranteed fallback correction for every non-trivial outcome, while zero-syndrome outcomes require no correction. Consequently, no outcome requires a coherent correction circuit more expensive than the original oracle. We connect these criteria to existing MBU techniques for modular arithmetic~\cite{Luongo2024MBU}, QROM~\cite{Gidney2019Windowed}, and QRAM routing~\cite{Cesa2025FastQRAM}, illustrating when
outcome-dependent compilation can simplify the correction further. Returning to the QEC perspective, we show that by choosing a different set of stabiliser checks and allowing for the redundant checks, we can further optimise the phase-coordinate representation of the correction operator to reduce its implementation cost.

The paper is organised as follows. \cref{sec:framework} develops the algebraic framework for post-selected
MBU, its QEC interpretation, and support discovery. \cref{sec:hhl-postselected-uncomputation} applies this framework to the HHL algorithm~\cite{Harrow_2009} to illustrate the potential for favourable resource scaling. We then turn to the phase-corrected MBU in \cref{sec:correctable_framework}, where the framework is generalised to accept non-trivial syndrome classes through feed-forward correction, encompassing both partially post-selected and fully deterministic regimes.  \cref{sec:conclusions} presents the conclusions and directions for future work. 

\section{Post-selected measurement-based uncomputation}
\label{sec:framework}
\subsection{Theoretical framework}
Let $F$ denote the main register and $G$ be an $N_g$-qubit garbage register. We consider the incoming state
\begin{equation}
 \ket{\Psi}_{FG}=\sum_{\lambda\in\Lambda}
 \alpha_\lambda\ket{\phi_\lambda}_F\ket{\lambda}_G,
 \label{eq:framework-state}
\end{equation}
where $\Lambda\subseteq\F_2^{N_g}$ is the populated garbage support, i.e. the set of garbage strings that have non-zero probability, and $\sum_\lambda|\alpha_\lambda|^2=1$. Our target state on the surviving main register is
\begin{equation}
 \ket{\Phi_0}=\sum_{\lambda\in\Lambda}\alpha_\lambda\ket{\phi_\lambda}.
 \label{eq:framework-target}
\end{equation}
We will refer to each term in the superposition as a \emph{branch}, labelled by its garbage string $\lambda$, and to the corresponding main-register states $\{\ket{\phi_\lambda}\}_{\lambda\in\Lambda}$ as the \emph{branch states}. Throughout the main text, we assume that the branch states are mutually orthonormal, i.e., $\braket{\phi_\mu}{\phi_\lambda}=\delta_{\mu\lambda}$. This is the natural setting when a single unitary clears the garbage while preserving every branch; the necessity of orthogonality for that mapping is discussed in \cref{sec:oracle condition}. Non-orthogonal branches are treated separately in \cref{app:nonorthogonal}.

Measurement-based uncomputation achieves this goal without coherently inverting the whole garbage-generation circuit (see \cref{figMain}(a)--(c)).
Instead, one measures the garbage register in the $X$ basis. By writing the $X$-basis states as
\begin{equation}
 \ket{k_X}=Z^k\ket{+}^{\otimes N_g},\qquad k\in\F_2^{N_g},
 \label{eqn:post_meas_state}
\end{equation}
where $Z^k=\bigotimes_i Z^{k_i}$, the post-measurement state on the main register, conditioned on outcome $k$ of the garbage register, is
\begin{align}
    (I_F\otimes \bra{k_X}_G)\ket{\Psi}_{FG}
&\propto
\ket{\Phi_k} = \sum_{\lambda\in \Lambda}(-1)^{k\cdot \lambda}\alpha_\lambda\ket{\phi_\lambda} \nonumber\\
& = (-1)^{k\cdot \lambda_0}\sum_{\lambda\in \Lambda}(-1)^{k\cdot (\lambda \oplus \lambda_0)}\alpha_\lambda\ket{\phi_\lambda}
\label{eq:framework-postmeasurement}
\end{align}
where $\lambda_0$ is an arbitrary but fixed element of $\Lambda$. Hence, the effect of the measurement does not destroy the coherent superposition, but introduces relative phases $(-1)^{k\cdot (\lambda \oplus \lambda_0)}$ between different terms.

In order to obtain the target state $\ket{\Phi_0}$ in \cref{eq:framework-target} \emph{up to a global phase}, we can post-select on outcomes that induce no relative phase between the terms, i.e., $k\cdot (\lambda \oplus \lambda_0) = 0$ for all $\lambda\in \Lambda$, giving rise to what we call \emph{post-selected MBU}. The accepted outcomes are defined by the set of constraints:
\begin{equation}
S_{\Lambda}
:=
\Span_{\F_2}\{\lambda\oplus \lambda_0:\lambda\in \Lambda\}
\subseteq \F_2^{N_g}.
\label{eq:framework-support}
\end{equation}
We call $S_\Lambda$ the \emph{affine support span} of the populated garbage support $\Lambda$.

To obtain the dimension of $S_{\Lambda}$, we need to look at two different cases. By putting $\lambda_0$ back into $S_{\Lambda}$, it can generate all elements in $\Span (\Lambda)$. Therefore, if $\lambda_0 \not \in S_{\Lambda}$, then $\lambda_0$ adds an independent dimension on top of $S_{\Lambda}$, while if $\lambda_0 \in S_{\Lambda}$ then adding back $\lambda_0$ adds no new dimension. Hence, we have 
\begin{align}
    r_{\Lambda} = \dim S_{\Lambda}  = \begin{cases}
     \dim (\Span (\Lambda)) - 1 \quad &\text{if } \lambda_0 \not \in S_{\Lambda}\\
    \dim (\Span (\Lambda)) \quad &\text{if } \lambda_0  \in S_{\Lambda}
    \end{cases}
    \label{eqn:affine_rank}
\end{align}
Here we use $r_{\Lambda}$ to denote the dimension of $S_{\Lambda}$, and we will call this the \emph{affine rank} of $\Lambda$, since it is the dimension of the affine support span of $\Lambda$. Note that we have $\lambda_0 \not \in S_{\Lambda}$ if and only if there does not exist any odd-sized subset of $\Lambda$ that sums to the zero vector, as shown in \cref{sec:proof_equiv_statement}. The definition of $S_{\Lambda}$ and whether $\lambda_0$ is contained in $S_{\Lambda}$ are entirely independent of our choice of $\lambda_0$. This will become clearer when we connect this to the quantum error correction perspective in the next subsection.

The set of all $k$ that we can post-select on to obtain our target state is given by the orthogonal complement of $S_{\Lambda}$:
\begin{equation}
S_{\Lambda}^\perp
:=
\{k\in \F_2^{N_g}:k\cdot s=0,\ \forall s\in S_{\Lambda}\},
\label{eq:framework-trivial}
\end{equation}
with $\dim S_{\Lambda}^\perp = N_g - \dim S_{\Lambda} = N_g - r_{\Lambda}$. 

Since the branch states are orthonormal, all possible measurement outcomes have the same probability of $2^{-N_g}$, and the success probability is proportional to the fraction of outcomes in $S_\Lambda^\perp$. This observation leads directly to the following theorem:

\begin{theorem}[Success probability of post-selected MBU]
\label{thm:succ-prob-post-selected-mbu}
Let $S_{\Lambda}$ be the affine support span of the populated garbage support $\Lambda$, with an arbitrarily chosen reference point $\lambda_0 \in \Lambda$, whose dimension is denoted as $r_{\Lambda} = \dim S_{\Lambda}$. Then the probability of obtaining an $X$-basis measurement outcome $k$ that leaves no relative phase on the populated support, equivalently $k\in S_\Lambda^\perp$, is

\begin{align*}
    P_{\mathrm{succ}} = 2^{-r_{\Lambda}} 
 = \begin{cases}
     2^{-\dim (\Span (\Lambda))+1} \quad &\text{if } \lambda_0 \not \in S_{\Lambda},\\
    2^{-\dim (\Span (\Lambda))} \quad &\text{if } \lambda_0  \in S_{\Lambda}.
    \end{cases}
\end{align*}
\end{theorem}

\cref{thm:succ-prob-post-selected-mbu} shows that the success probability of post-selected MBU depends only on the affine rank $r_\Lambda$ of the populated garbage support $\Lambda$, rather than on the total number $N_g$ of garbage qubits. Consequently, whenever its affine support span $S_{\Lambda}$ occupies a low-dimensional affine subspace of $\mathbb{F}_2^{N_g}$, post-selected MBU remains efficient despite the presence of a large garbage register. In terms of implementation, post-selected MBU replaces the entire uncomputation unitary with the X-basis measurements on the garbage register, at the cost of a sampling overhead of $P_{\mathrm{succ}}^{-1} = 2^{r_{\Lambda}}$. We will extend this framework in \cref{sec:correctable_framework} by correcting the non-trivial phases via feed-forward operations, including the partially post-selected and deterministic MBU.

\subsection{QEC perspective}
\label{subsec:qec-perspective}
The algebraic framework developed above in \cref{thm:succ-prob-post-selected-mbu} admits a natural interpretation in terms of quantum error correction, illustrated schematically in \cref{figMain}(d). Here, using \cref{eqn:post_meas_state}, we can view the process of measuring out the garbage register as inflicting $Z^k$ errors on the register when we obtain a different measurement outcome $k$. As shown in \cref{eq:framework-postmeasurement}, we want to post-select on states with errors $k$ that satisfy $k\cdot s = 0$ for all $s \in S_{\Lambda}$, where $(-1)^{k\cdot s}$ is the \emph{relative phase} between different branches in the superposition. Hence, if we view $S_{\Lambda}$ as the $X$-stabilisers, the relative phase pattern $\{k\cdot s: s \in S_{\Lambda}\}$ is simply the (over-complete) set of error syndromes we get when we measure the $X$-stabilisers on the state after the $Z^k$ error. The condition that $k\cdot s = 0$ for all $s \in S_{\Lambda}$ is then equivalent to post-selecting on the syndrome-zero subspace of the code defined by the stabilisers $S_{\Lambda}$. This means post-selecting on $k$ that corresponds to $Z$-stabilisers and the logical $\overline{Z}$, which are elements in $S_{\Lambda}^\perp$ and commute with the $X$-stabilisers. 

To fully specify the code, we can also define its $Z$-stabilisers with their support being $S_Z = \Span(\Lambda)^{\perp}$, which naturally commute with the $X$ stabilisers. When $\lambda_0 \in S_{\Lambda}$, all elements in $\Lambda$ maps to the $X$-stabilisers, which means we have $k \cdot \lambda = 0$ for all $\lambda \in \Lambda$ since all $Z$-stabilisers commute with all $X$-stabilisers. In this case, there are no $\overline{Z}$ and $\overline{X}$ logical operators since $\dim S_\Lambda + \dim S_Z= N_g$, so there is no logical degree of freedom left. When $\lambda_0 \notin S_{\Lambda}$, there is one logical degree of freedom left; we can view $\lambda_0$ as a logical $\overline{X}$, which is naturally independent of the stabilisers $S_{\Lambda}$. In this case, elements in $\Lambda$ are no longer stabilisers; they are some representative of the logical $\overline{X}$. For all elements $\lambda \in \Lambda$, we will obtain phase $k \cdot \lambda = 0$ for $k$ that correspond to $Z$-stabilisers and phase $k \cdot \lambda = 1$ for $k$ that correspond to a logical $\overline{Z}$. Note that when we talk about stabilisers and syndrome from now on, by default we are talking about $X$ stabilisers and the syndrome from these $X$ checks since we only care about $Z$ errors.

Post-selecting on the syndrome-zero subspace $S_{\Lambda}^\perp$ is the most straightforward way to ensure that there is no measurement-induced relative phase between different superposed terms in the output state. In some application scenarios, even if we obtain non-trivial syndromes that induce non-trivial relative phase patterns, we can still accept the measurement outcomes as long as the relative phases can be removed by relatively simple operations (compared to full uncomputation) on the main register. We will defer the discussion of such cases to \cref{sec:correctable_framework}. As mentioned, different relative phase patterns are induced by different error syndromes. Therefore, the number of possible relative phase patterns is simply the number of syndrome subspaces for $X$ checks, which is exponential in the number of $X$ stabiliser checks, i.e.,  $2^{\dim(S_{\Lambda})} = 2^{r_\Lambda}$.

\begin{figure*}
\includegraphics[width=0.9\textwidth]{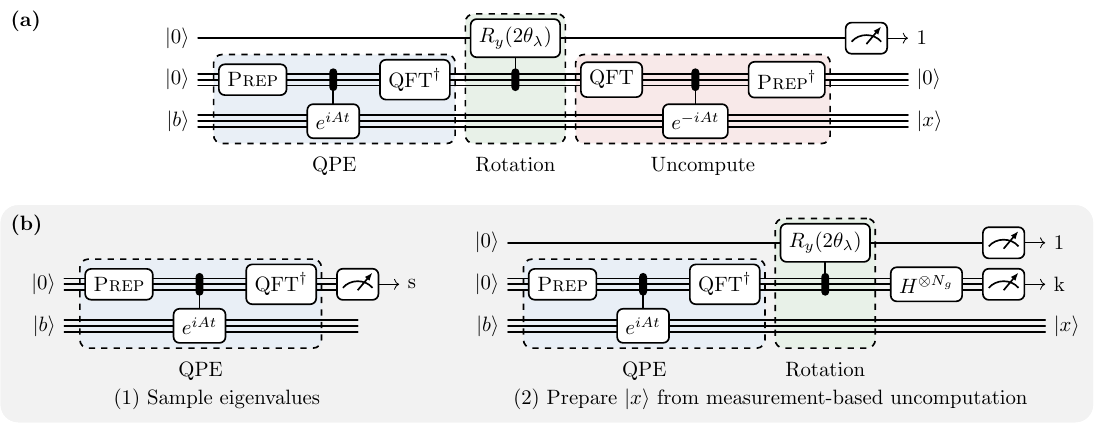}
\caption{\emph{Comparison between coherent and post-selected uncomputation in HHL.} (a) Standard coherent HHL implementation. After QPE computes the eigenvalue estimate into the phase register and the controlled rotation implements the factor proportional to $1/\lambda$, the eigenvalue register is disentangled by applying inverse QPE. This requires the phase-estimation workspace to remain coherent until the end of the inverse block. (b) Post-selected measurement-based replacement. A preliminary sampling stage estimates the populated eigenvalue support $\Lambda$ and hence the affine support span $S_{\Lambda}$. In the state-preparation run, the controlled rotation is followed by an $X$-basis measurement of the eigenvalue register. The measurement outcome $k$ is accepted when $k\in S_{\Lambda}^\perp$, in which case the induced phase is constant on the populated eigenvalue support and no inverse QPE is required. Accepted runs therefore prepare the desired HHL output state, up to the usual ancilla post-selection for the controlled rotation and an irrelevant global phase, and up to the support-truncation error when the estimated support is incomplete (see \cref{subsubsec:support-discovery}).} 
\label{fig:hhl-mbu}
\end{figure*}

\subsection{Support discovery and sample complexity}
\label{subsubsec:support-discovery}
Implementing this pure post-selection requires prior knowledge of the populated support of the set of garbage strings $\Lambda$ (note that we only need to obtain its generators instead of the full set). Instead of trying to obtain the exact target state $\ket{\Phi_0}$ in \cref{eq:framework-target}, we can set a resolution threshold $\mu > 0$, such that we only want to faithfully recover the components that have output probability above the threshold $p_\lambda = |\alpha_\lambda|^2 \ge \mu$. By measuring the garbage register of our starting state before uncomputation (\cref{eq:framework-state}) and following the probability threshold $\mu$, we can efficiently approximate the support $\Lambda$ using a classical heavy-hitters sampling protocol~\citep{motwani1995randomized,mitzenmacher2005probability}. As detailed in \cref{app:sec:support-discovery}, achieving this with a failure probability of at most $\delta_{\mathrm{fail}}$ requires a sample complexity of $N_{\mathrm{samp}} = \mathcal{O}\left(\frac{1}{\mu} \log \frac{1}{\mu\delta_{\mathrm{fail}}}\right)$. Any rare branches with $p_\lambda < \mu$ that evade detection constitute a statistical tail mass, $\epsilon_{\rm tail}$, which translates directly to an $\mathcal{O}(\sqrt{\epsilon_{\rm tail}})$ penalty in the $\ell_2$ state-vector norm of the final uncomputed state. 

Instead of sampling from the starting state before uncomputation (\cref{eq:framework-state}), we may be able to prepare another state at a cheaper cost that still has the same support of the garbage strings, e.g., a state of the form $\ket{\Psi_{\Lambda}}  =  \sum_{\lambda \in \Lambda} \beta_\lambda \ket{u_\lambda} \ket{\lambda}$. Estimating the support $\Lambda$ by measuring the garbage register of this new state will reduce the complexity of the quantum circuit required, at the cost of possibly increasing the sample complexity of the support estimation. Measuring the garbage register on this new state is sampling from the probability distribution $\{|\beta_\lambda|^2\}$, but our threshold is still defined on the probability distribution $\{|\alpha_\lambda|^2\}$. Let $\kappa^2$ be the maximum amplification factor of the probabilities from $\{|\beta_\lambda|^2\}$ to $\{|\alpha_\lambda|^2\}$, i.e., $\kappa^2 = \max_{\lambda} \frac{|\alpha_\lambda|^2}{|\beta_\lambda|^2}$. Then we can still use the same heavy-hitters sampling protocol by applying a tightened initial threshold of $|\beta_\lambda|^2 \ge \mu/\kappa^2$. The sample complexity is then given by $N_{\mathrm{samp}} = \mathcal{O}\left(\frac{\kappa^{2}}{\mu} \log \frac{1}{\mu\delta_{\mathrm{fail}}}\right)$. 

\section{Case study: Affine rank scaling in the HHL algorithm}
\label{sec:hhl-postselected-uncomputation}
\subsection{Mapping to post-selected uncomputation}
In this section, we will apply post-selected uncomputation to the Harrow-Hassidim-Lloyd (HHL) algorithm~\citep{Harrow_2009} to replace the inverse quantum phase estimation (QPE)~\citep{kitaev1995quantum,nielsen2010quantum,cleve1998quantum} block necessary for uncomputing the eigenvalue register. The post-selection sampling overhead is set by the affine rank $r_{\Lambda}$ through \cref{thm:succ-prob-post-selected-mbu}, and we show that for some very structured problems this affine rank can be bounded, so that post-selected MBU incurs only a constant-factor sampling overhead that is independent of the system size at fixed input bandwidth and eigenvalue-register precision.

 In the HHL algorithm, we are given a Hermitian matrix $A$ and a vector $\ket{b}$, and we want to solve the linear system $A\ket{x} \propto \ket{b}$. They can be written in terms of the eigenbasis of $A$ as:
    \begin{align*}
        A &= \sum_{\lambda \in \Lambda} \lambda \Pi_\lambda\\
        \ket{b} &= \sum_{\lambda \in \Lambda}\beta_\lambda\ket{\phi_\lambda}\\
        \ket{x} &= \widetilde{C} \sum_{\lambda \in \Lambda} \frac{\beta_\lambda}{\lambda}\ket{\phi_\lambda}
    \end{align*}
    where $\widetilde{C}$ is a normalising constant.

The algorithm consists of three main steps with the circuit shown in \cref{fig:hhl-mbu}(a): 
\begin{enumerate}
    \item Quantum phase estimation (QPE) to encode the eigenvalues of $A$ into an ancillary register,
    \item A controlled rotation and post-selection to apply the reciprocal of the eigenvalues, and
    \item Inverse QPE to uncompute the eigenvalue register.
\end{enumerate}
Right after step (2) and before the uncomputation using inverse QPE, the state of the system is a superposition of the form
 \begin{align*}
    \ket{\Psi}_{FG} &=  \sum_{\lambda \in \Lambda} \widetilde{C}\frac{\beta_\lambda}{\lambda}\ket{\phi_\lambda}_F\ket{\lambda}_G.
\end{align*}
For simplicity, here we assume perfect QPE, such that the eigenvalue register stores a sufficient $t$-bit binary representation $\lambda$ of the eigenvalues. All implementation details arising from finite precision and tapering are deferred to \cref{app:spectral-hhl}. We see that this is the same as \cref{eq:framework-state} with $\alpha_\lambda = \widetilde C \frac{\beta_\lambda}{\lambda}$ and our target state $\ket{x}$ is the same as \cref{eq:framework-target}. Hence, we can apply the post-selected MBU framework to this state to uncompute the eigenvalue register. By measuring the eigenvalue register in the $X$ basis, we can post-select on the outcomes $k\in S_{\Lambda}^\perp$ to obtain the target state $\ket{x}$ up to a global phase. The success probability of this post-selection is given by \cref{thm:succ-prob-post-selected-mbu}, which is determined by $r_\Lambda$, the affine rank of the eigenvalue support $\Lambda$.

\cref{fig:hhl-mbu} illustrates the replacement of the coherent inverse-QPE block in standard HHL with post-selected MBU, i.e., an $X$-basis measurement of the eigenvalue register followed by a post-selection on the syndrome-zero subspace, i.e., on the outcomes $k\in S_{\Lambda}^{\perp}$.

\subsection{Affine rank scaling}
\label{sec:numerics}
 By \cref{thm:succ-prob-post-selected-mbu}, the sampling overhead of post-selected MBU is $2^{r_{\Lambda}}$, so the cost of the case study is controlled entirely by how the affine rank of the populated eigenvalue support scales. For highly structured systems, such as Laplacians of highly symmetric graphs, discretised differential equations, or any operator diagonalisable in a Fourier-like basis $A = F^{\dagger}\,\mathrm{diag}(\lambda_{1},\lambda_{2},\dots)\,F$, smooth input states concentrate on a small number $K$ of low-frequency modes. When the discretised labels of these populated eigenvalues are algebraically structured, they can, in some cases, lie on a low-dimensional subset of $\F_{2}^{N_{g}}$, leading to an affine rank as low as $r_{\Lambda}=\mathcal{O}(\log_{2}K)$. Note that this is due to the additional properties from the structured spectral-encoding. In this regime, the exponential sampling overhead collapses to a constant factor independent of the system size at fixed bandwidth and precision, while each accepted shot still enjoys the benefits of MBU: the per-shot depth is roughly halved when the QPE blocks dominate the circuit depth, and the $N_{g}$ eigenvalue-register qubits are released immediately upon measurement. The explicit matrix structure requirements and the extended dequantisation arguments are given in \cref{app:subsec:hhl-structure,app:subsec:dequantization}.

\begin{figure}
\centering
{\includegraphics[width=\linewidth]{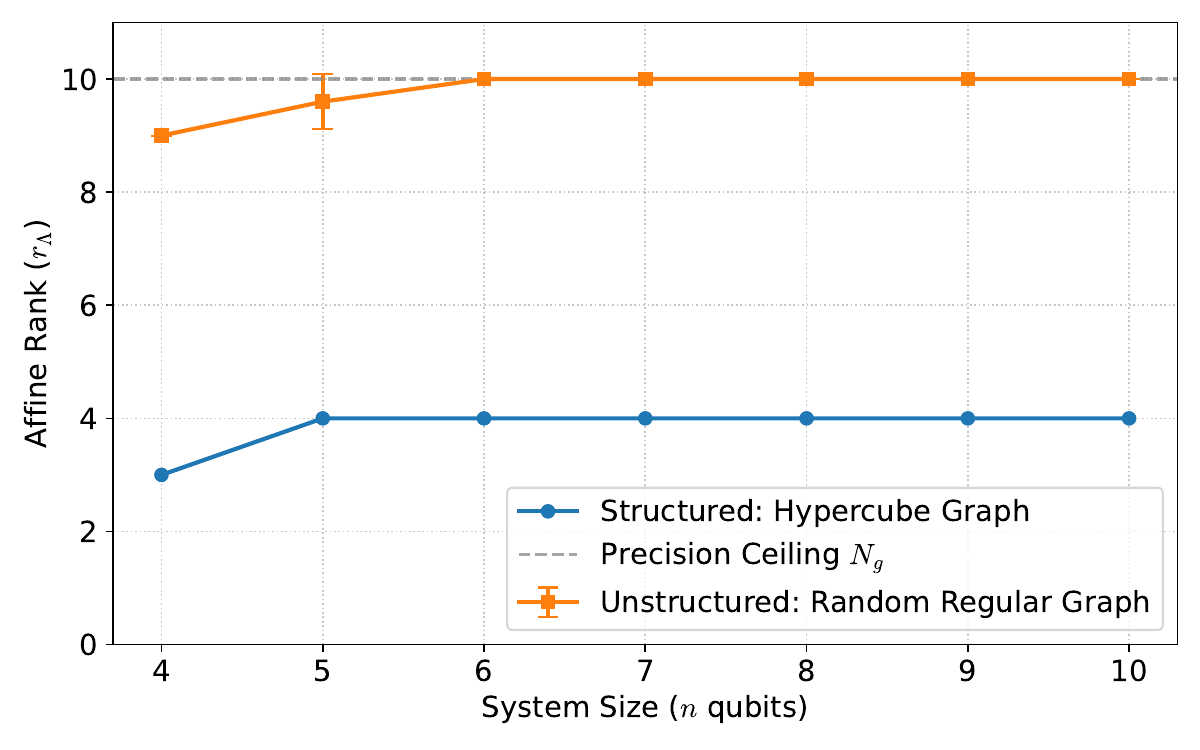}}
\caption{\emph{Numerical validation of MBU for structured graph Laplacians.} Affine rank $r_\Lambda$ as a function of system size $n$ for a smooth physical signal (exponential spatial decay). For the structured hypercube graph, the affine rank remains functionally decoupled from the exponentially growing system volume over the simulated range. In contrast, eigenvalues of an unstructured random regular graph (RRG) lack algebraic correlation, causing the discretised labels to behave as independent bitstrings and $r_\Lambda$  to rapidly saturate the precision ceiling $N_{g}=10$. For the random regular graph, the affine rank is averaged over 10
independent graph instances with error bars denoting one standard deviation. Since $r_\Lambda\le N_g$ by construction, the contrast shown is over the tested range and does not by itself establish asymptotic scaling; at fixed algorithmic accuracy, $N_g$ must grow with the required eigenvalue precision.}
\label{fig: r_lambda_vs_n}
\end{figure}

We validate this scaling numerically on the Laplacian of an $n$-dimensional hypercube graph (with $N=2^{n}$ vertices), a topology highlighted as a favourable candidate for quantum linear solvers~\cite{Shetty_2026}. The input states $\ket{b}$ model a smooth, localised physical signal whose amplitudes decay exponentially with the shortest-path graph distance (Hamming weight) from a source node, $b_{x}\propto e^{-\gamma d(x,x_{0})}$. Note that $k=0$ mode corresponds to the kernel of the hypercube graph Laplacian, and any component of $\ket{b}$ is removed by the controlled-rotation postselection as in the standard HHL (see \cref{app:subsec:hypercube-theory} for more details). For each system size we diagonalise the Laplacian exactly (using the NetworkX~\citep{Hagberg2008NetworkX} and NumPy~\citep{Harris2020NumPy} packages), retain the eigenvalues whose overlap with $\ket{b}$ exceeds a fixed threshold as the populated support, round these eigenvalues to a fixed-precision binary label to model the finite eigenvalue register produced by phase estimation, and compute the affine rank of the resulting labels over $\F_{2}$. For this product-form input the eigenspace weights are binomial, so at fixed $\gamma$ the effective bandwidth grows with $n$; over the tested range the omitted probability mass remains at or below the per-cent level, and the full bandwidth analysis, including the smoothness scaling under which the bandwidth stays bounded, is given in \cref{app:subsec:hypercube-theory}. As shown in \cref{fig: r_lambda_vs_n}, the affine rank of the hypercube graph remains functionally decoupled from the exponentially growing system volume, because the populated labels lie on a low-dimensional subspace of $\F^{N_{g}}_{2}$ and do not populate new independent degrees of freedom over the tested range. In contrast, the discretised eigenvalues of an unstructured random regular graph of identical degree and node count lack algebraic correlation and behave as independent bitstrings, causing the affine rank to rapidly saturate the precision ceiling $N_{g}$. \cref{fig: r_lambda_vs_K} in \cref{app:subsec:hypercube-theory} isolates the dependence on the input bandwidth $K$ and confirms the expected $\mathcal{O}(\log_{2}K)$ scaling, together with the spectral analysis of the hypercube graph. The resulting space--time trade-offs on early fault-tolerant devices, including a worked example with concrete repetition counts and code-distance savings, are given in \cref{sec:EFT_trade_off}.

\section{Phase-corrected measurement-based uncomputation}
\label{sec:correctable_framework}

\cref{sec:framework,sec:hhl-postselected-uncomputation} focused on the pure post-selection regime of MBU, where only the syndrome-zero outcomes are accepted. As shown in \cref{sec:EFT_trade_off}, the sampling overhead can remain significant even when it is independent of the size of the system. It is thus natural to ask whether it is possible to accept more (or even all) measurement outcomes beyond the zero syndrome and remove the non-trivial relative phase patterns induced by MBU by applying additional operations on the main register. In this section, we develop this generalisation and identify the conditions under which different measurement outcomes can be corrected using the original uncomputation oracle and beyond.

\subsection{Phase correction using the uncomputation oracle}
\label{subsec:deterministic-criterion}

Recall from \cref{sec:framework} that measuring the garbage register in the $X$-basis introduces an outcome-dependent relative phase $(-1)^{k\cdot\lambda}$ between the superposed branch states. Here, we will look at under what conditions we can directly use the uncomputation oracle to correct the remaining phases in \cref{eq:framework-postmeasurement} resulting from the garbage measurement, making the MBU deterministic.

We will use $U_\phi$ to denote the uncomputation unitary that achieves the uncomputation mapping from the pre-uncomputation state $\ket{\Psi}_{FG}$ in \cref{eq:framework-state} to the target state in $\ket{\Phi_0}$ in \cref{eq:framework-target}. One natural form of such an oracle that can also be used to correct the phase induced by the garbage measurement is given as below.
\begin{theorem}[MBU phase correction from the structure of the uncomputation oracle]
\label{thm:branch-controlled-deterministic}
Let $U_{\phi}$ be a unitary uncomputation oracle whose action on the target state branch subspace, represented by the projection operator $\Pi_\Lambda=\sum_{\lambda\in\Lambda}\ketbra{\phi_\lambda}$, takes the block form of:
\begin{equation}
U_\phi\Pi_\Lambda
=
\sum_{\lambda\in\Lambda}
\ketbra{\phi_\lambda}\otimes X^{\lambda},
\label{eq:block-form-independent}
\end{equation}
or more generally, 
\begin{equation}
U_\phi\Pi_\Lambda
=
(U_{s,F} \otimes W) \sum_{\lambda\in\Lambda}
\ketbra{\phi_\lambda}\otimes X^{\lambda},
\label{eq:block-form-independent_gen}
\end{equation}
for some unitary $U_{s,F}$ acting on the main register and stabilises the target state $U_{s,F} \ket{\Phi_0} = \ket{\Phi_0}$ while $W$ can be any unitary acting on the garbage register. Then measuring the garbage register in the $X$ basis with any outcome and applying the same oracle $U_{\phi}$ yields the state $\ket{\Phi_0}$ on the main register.
\end{theorem}
This can be proved by directly acting $U_\phi$ on the post-measurement state in \cref{eq:framework-postmeasurement}. Even though it seems that \cref{eq:block-form-independent_gen} takes quite a restrictive and special form, it is actually the only possible uncomputation oracle structure that allows for phase correction for all outcomes if we assume the oracle can uncompute individual branch, and its actions on the garbage register in individual branches are all unitary as shown in \cref{sec:oracle condition}. Note that \cref{eq:block-form-independent} only specifies $U_\phi$ on the subspace defined by $\Pi_\Lambda$ on the main register; its action on the orthogonal complement can be chosen arbitrarily, subject to unitarity. 

\textbf{Example: Modular Arithmetic. } The simplest example is the modular arithmetic example of Ref.~\cite{Luongo2024MBU}, in which the garbage is a single comparator qubit storing a Boolean predicate $g(x)$ of the data $x$ in the surviving arithmetic registers. The uncomputation oracle is the branch-controlled bit flip $U_{\phi}=\sum_{x}\ketbra{x}\otimes X^{g(x)}$, which is exactly the block form of \cref{eq:block-form-independent}. \cref{thm:branch-controlled-deterministic} therefore applies directly: the syndrome-zero outcome $k=0$ requires no correction, while for $k=1$ applying the same comparator oracle cancels the measurement-induced phase, so every outcome is accepted and the MBU is deterministic. The detailed construction is given in \cref{app:deterministic-modular}.

In contrast, in the HHL example of \cref{sec:hhl-postselected-uncomputation} the inverse QPE block for uncomputation does not fit \cref{eq:block-form-independent_gen} since its branch-$\lambda$-dependent action on the garbage state is some $W_\lambda X^{\lambda}$ rather than $X^\lambda$. Applying the same oracle after the measurement cancels the measurement-induced phases as in \cref{eq:auto-phase-cancellation}, but it leaves the main register entangled with the branch-dependent residual garbage states $W_{\lambda}\ket{k_X}$, so the target state $\ket{\Phi_0}$ is not recovered. An explicit derivation is given in \cref{app:hhl-deterministic-failure}. This is a limitation of reusing uncomputation oracles for phase correction, not a fundamental obstruction to constructing an alternative tailored phase correction oracle. \cref{sec:hhl-postselected-uncomputation} uses pure syndrome-zero post-selection to isolate the uncomputation overhead without assuming an additional correction circuit.

\subsection{Adapting the uncomputation oracle to the measurement outcome}
\label{subsec:deterministic-examples}
The phase correction of \cref{thm:branch-controlled-deterministic} can also be understood through phase kickback. Decomposing the bit-flip part of the oracle in the measurement basis, $X^{\lambda}=\sum_{k}(-1)^{k\cdot\lambda}\ketbra{k_X}$, the block form of \cref{eq:block-form-independent} factorises as
\begin{equation}
U_{\phi}\Pi_{\Lambda}
= \sum_{k}D_{k}\otimes\ketbra{k_X},
\label{eq:deferred-measurement-form}
\end{equation}
where $D_k$ is the phase correction operator for the main register corresponding to the measurement outcome $k$, whose action on the subspace spanned by the branch states is given by
\begin{align}
D_k \Pi_\Lambda =&\sum_{\lambda\in\Lambda}(-1)^{k\cdot\lambda}\ketbra{\phi_\lambda}.
\label{eq:main-register-diagonal}
\end{align}
In this picture, it is easier to see that since $X$-basis measurement commute with the $X$-basis controlled component of the uncomputation oracle, we can insert $X$ basis measurement immediately before the uncomputation oracle without changing the overall effect. In fact, the uncomputation oracle is just one way to effectively implement $D_k$ on the main register conditioned on the measurement outcome $k$. Instead of applying the same uncomputation oracle $U_{\phi}$ for all measurement outcomes, we can adapt the uncomputation oracle to the measurement outcome $k$ and implement a simpler oracle $U_k$ instead. In some cases, we may be able to implement $D_k$ directly on the main register without any additional ancilla at a cheaper cost than those requiring ancilla (including the full uncomputation oracle).

\textbf{Example: QROM lookup. }The QROM unlookup of Ref.~\cite{Gidney2019Windowed} illustrates this adaptation. The garbage stores the table values $T(a)$ indexed by the data $a$ in the surviving address register, and the unlookup oracle is the branch-controlled bit flip $U_{\phi}=\sum_{a}\ketbra{a}\otimes X^{T(a)}$, i.e., the block form of \cref{eq:block-form-independent}, so \cref{thm:branch-controlled-deterministic} already guarantees deterministic MBU with the original unlookup, at the $\mathcal{O}(L)$ Toffoli cost of the full coherent unlookup for a table of size $L$. Since the lookup table is available at compile time and the outcome $k$ is classically known, with the help of an ancillary unary register, we can instead implement an outcome-adapted oracle $U_k$ compiled from reduced fixup tables, whose Toffoli cost scales as $\mathcal{O}(\sqrt{L})$. The detailed construction is given in \cref{app:deterministic-lookup}. 

\textbf{Example: QRAM routing. }The QRAM routing example of Ref.~\cite{Cesa2025FastQRAM} provides another example of such an outcome-adapted oracle. The encoding map $U_{\mathrm{NOHE}}$ is an isometry that maps the address register into a larger physical register, meaning its inverse only needs to be defined on the image of the encoding; introducing ancilla qubits initialised in $\ket{0}$ realises the encoding as a unitary dilation $\widetilde{U}$ and reduces the problem to the unitary setting of \cref{subsec:deterministic-criterion}. The dilated inverse splits into a branch-independent Clifford part, applied coherently, and a Toffoli network $\sum_{i,j}\ketbra{i,j}\otimes X^{i\wedge j}$. After the fixed Clifford preprocessing, distinct pairs $(i,j)$ with the same garbage value $\lambda=i\wedge j$ are grouped coherently into one
normalised branch state $\ket{\phi_\lambda}$, thus the remaining Toffoli factor is in the block form of \cref{eq:block-form-independent}. After the target qubit is measured in the $X$ basis with outcome $k$, the outcome-adapted correction on the surviving address qubits is simply $D_k = \mathrm{CZ}^{k}$, and this feed-forward correction is cleanly implemented through an adaptive choice of measurement basis together with Pauli-frame updates. The detailed construction is given in \cref{app:deterministic-qram}.

\subsection{Optimising the phase-correction circuit}
\subsubsection{Minimal phase correction bases}
The optimisation over the outcome-dependent recovery operations $U_k$ and $D_k$ concerns the circuit used to realise a correction. Independently of that optimisation, the
phase-correction problem itself admits an exact algebraic compression. Although the raw garbage label $\lambda$ and the raw measurement outcome $k$ are both $N_g$-bit strings, their branch-dependent pairing is supported only on the $r_{\Lambda}$-dimensional affine support span $S_{\Lambda}$. This reduces the effective correction problem from $\mathbb{F}_{2}^{N_g}$ to $\mathbb{F}_{2}^{r_{\Lambda}}$.

Recall from \cref{subsec:qec-perspective} that $S_{\Lambda}$ defined from the garbage support can be seen as a set of stabilisers. We can define a set of $r_\Lambda$ stabiliser generators $\{\widetilde{s}_1, ..., \widetilde{s}_{r_\Lambda}\}$ for $S_{\Lambda}$ which give rise to the stabiliser generator matrix:
\begin{align} \label{eqn:gen_matrix}
    \widetilde{\boldsymbol{\mathrm{S}}}_\Lambda := \begin{pmatrix}
        \widetilde{s}_1^T\\
        \vdots\\
        \widetilde{s}_{r_\Lambda}^T
    \end{pmatrix} \in \mathbb{F}_{2}^{r_{\Lambda} \times N_g}
\end{align}
For a given measurement outcome $k$, as discussed in \cref{subsec:qec-perspective}, we can view them as errors which will give rise to error syndromes that correspond to the induced relative phase between different branches in the post-measurement state. By switching from the overcomplete set of stabilisers in $S_{\Lambda}$ to the stabiliser generators $\widetilde{\boldsymbol{\mathrm{S}}}_\Lambda$, we can obtain a compressed error syndrome $q(k)$ that is more akin to the regular error syndrome we encounter in regular QEC:
\begin{align}\label{eqn:syndrome_vec}
    q(k) := \widetilde{\boldsymbol{\mathrm{S}}}_\Lambda k \in \mathbb{F}_{2}^{r_{\Lambda}}. 
\end{align}
Different $q(k)$ correspond to the different syndrome subspace, and the space of $q(k)$ is a quotient space defined by $Q=
\mathbb{F}^{N_g}_2 / S_{\Lambda}^{\perp}$ with $S_{\Lambda}^{\perp} = \mathrm{Ker} \widetilde{\boldsymbol{\mathrm{S}}}_\Lambda$.

For a given stabiliser $\lambda \oplus \lambda_0 \in S_\Lambda$, we can always decompose it into the stabiliser generators:
\begin{align}\label{eqn:gen_rep_vec}
    \lambda \oplus \lambda_0 = \sum_{i = 1}^{r_\Lambda} \gamma_i(\lambda) \widetilde{s}_i = \widetilde{\boldsymbol{\mathrm{S}}}_\Lambda^{T} \gamma(\lambda)
\end{align}
where $\gamma(\lambda)$ is the vector representation of the stabiliser in the stabiliser generator basis. It is unique and can be found by solving the linear system above. 

In this way, together with \cref{eqn:syndrome_vec}, the exponent of the induced phase can be re-expressed as:
\begin{align*}
    k \cdot (\lambda \oplus \lambda_0) = k^T \widetilde{\boldsymbol{\mathrm{S}}}_\Lambda^{T} \gamma(\lambda) = q(k) \cdot \gamma(\lambda). 
\end{align*}
The post-measurement states in \cref{eq:framework-postmeasurement} can therefore be written as:
\begin{align}
\ket{\Phi_k} = (-1)^{k\cdot \lambda_0}\sum_{\lambda\in \Lambda}(-1)^{q(k) \cdot \gamma(\lambda)}\alpha_\lambda\ket{\phi_\lambda}
\label{eqn:new_post_meas_state}
\end{align}

\subsubsection{Further phase function transformation}
So far the representation we consider is minimal in dimension, since both $q(k)$ and $\gamma(\lambda)$ contain only $r_\Lambda$ bits. However, a minimal-dimensional
representation does not necessarily give the lowest-cost implementation of the phase correction. We can introduce redundant phase coordinates which leave the represented phase unchanged but provide the compiler with additional choices of correction circuits.

We can further generalise the minimal generator-coordinate
representation by allowing an overcomplete stabiliser representation.
Let
\begin{equation}
\boldsymbol{\mathrm{H}}_\Lambda\in\F_2^{m\times N_g},
\qquad m\geq r_\Lambda,
\end{equation}
whose rows form a spanning set of the stabiliser space $S_\Lambda$ and thus $\operatorname{rank}(\boldsymbol{\mathrm{H}}_\Lambda)=r_\Lambda$. Hence, for each stabiliser $\lambda\oplus\lambda_0 \in S_\Lambda$, we can always solve for a
coefficient vector $b(\lambda)\in\F_2^m$ satisfying
\begin{equation}
\boldsymbol{\mathrm{H}}_\Lambda^T b(\lambda)
=
\lambda\oplus\lambda_0.
\label{eq:valid-overcomplete-coordinate}
\end{equation}
Let $\boldsymbol{\mathrm K}_\Lambda$ be a fixed linear right inverse of
$\boldsymbol{\mathrm H}_\Lambda^T$ on $S_\Lambda$ (such a right inverse
exists because the image of $\boldsymbol{\mathrm H}_\Lambda^T$ is
$S_\Lambda$):
\begin{align*}
    \boldsymbol{\mathrm H}_\Lambda^T\boldsymbol{\mathrm K}_\Lambda (\lambda\oplus\lambda_0)
 &=\lambda\oplus\lambda_0 \quad\forall \lambda \in \Lambda,
\end{align*}
then the general form of $b(\lambda)$ is given by:
\begin{equation}
 b(\lambda)
 =\boldsymbol{\mathrm K}_\Lambda(\lambda\oplus\lambda_0)
  \oplus z(\lambda),
 \label{eq:general-phase-representation}
\end{equation}
where the function $z$ is any function that map $\Lambda$ into the kernel of $\boldsymbol{\mathrm H}_\Lambda^T$, i.e. $z:\Lambda\longrightarrow\ker\boldsymbol{\mathrm H}_\Lambda^T$, and thus can be non-linear. 

For a measurement outcome $k$, we can obtain the corresponding
overcomplete syndrome representation
\begin{equation}
a(k):=\boldsymbol{\mathrm{H}}_\Lambda k\in\F_2^m.
\label{eq:overcomplete-syndrome}
\end{equation}
Then, for every $\lambda\in\Lambda$,
\begin{align}
k\cdot(\lambda\oplus\lambda_0) = k^T \boldsymbol{\mathrm{H}}_\Lambda^T b(\lambda)
 = a(k)\cdot b(\lambda)
\label{eq:overcomplete-phase-factorisation}
\end{align}
Hence the post-measurement state may equivalently be written as
\begin{equation}
\ket{\Phi_k}
=
(-1)^{k\cdot\lambda_0}
\sum_{\lambda\in\Lambda}
(-1)^{a(k)\cdot b(\lambda)}
\alpha_\lambda\ket{\phi_\lambda}.
\end{equation}
Follows the exact same argument as \cref{subsec:deterministic-criterion}, if we can find an oracle that takes the form of \cref{eq:block-form-independent} with $b(\lambda)$ in place of $\lambda$, we can then prepare ancilla in the state of $Z^{a(k)} \ket{+}$ based on the measurement outcome $k$ and then apply this oracle to get the desired $(-1)^{a(k) \cdot b(\lambda)}$ phase correction. 

In QEC terms, the rows of $\boldsymbol{\mathrm{H}}_\Lambda$ specify a possibly redundant set of $X$-checks. The vector $a(k)$ records their syndromes, while $b(\lambda)$ selects checks whose product is $X^{\lambda\oplus\lambda_0}$. This change of representation allows a flexible design choice for the phase-correction circuit. Rather than implementing the original uncomputation oracle, which computes the map $\phi_{\lambda}\mapsto\lambda$ across the full $N_g$-qubit garbage register, the compiler only needs to construct a circuit that computes the transformed map $\phi_{\lambda} \mapsto b(\lambda)$, which can be cheaper to implement. The transformation replaces the original garbage label with new representation $b(\lambda)$, which may reduce the number of ancilla qubits required for phase correction when \(m<N_g\), but a larger redundant representation may also be useful if it simplifies the correction circuit as will be discussed in a later unary-representation-assisted QROM example. More importantly, because the transformation matrix $H_{\Lambda}$ can be freely chosen as long as its row space spanned the stabiliser space $S_\Lambda$, the representation $b(\lambda)$ can be explicitly co-optimised alongside the correction circuit to minimise the overall gate cost. 

\textbf{Example: copying a linear function. }
The simplest instance occurs when the compute--copy--uncompute pattern~\cite{bennett1973logical} copies a linear function of the data, with CNOT fan-outs or parity-check ancillas acting as the simplest instances. The branch states are computational-basis states $\ket{\phi_\lambda}$ of $n$ data qubits and the garbage string is $\lambda=\boldsymbol{\mathrm M}\phi_\lambda$, where $\boldsymbol{\mathrm M}\in\F_2^{N_g\times n}$. We first consider the case in which the populated support is sufficiently rich that $S_\Lambda=\operatorname{Im}\boldsymbol{\mathrm M}$. In this case, we may choose the overcomplete stabiliser representation $\boldsymbol{\mathrm H}_\Lambda=\boldsymbol{\mathrm M}^T$, whose row space is exactly $S_\Lambda$.
Choosing a reference branch $\phi_{\lambda_0}$ with $\lambda_0=\boldsymbol{\mathrm M}\phi_{\lambda_0}$, the representation condition $\boldsymbol{\mathrm H}_\Lambda^T b(\lambda)=\lambda\oplus\lambda_0$ reduces to $\boldsymbol{\mathrm M}b(\lambda)=\boldsymbol{\mathrm M}(\phi_\lambda\oplus\phi_{\lambda_0})$, which is solved by $b(\lambda)=\phi_\lambda\oplus\phi_{\lambda_0}$. The corresponding syndrome representation is $a(k)=\boldsymbol{\mathrm H}_\Lambda k=\boldsymbol{\mathrm M}^Tk$. Hence, the phase correction reduces to applying $Z^{\boldsymbol{\mathrm M}^Tk}$ directly to the main register, since $Z^{\boldsymbol{\mathrm M}^Tk}\ket{\phi_\lambda}=(-1)^{k\cdot\boldsymbol{\mathrm M}\phi_\lambda}\ket{\phi_\lambda}=(-1)^{k\cdot\lambda}\ket{\phi_\lambda}$, exactly cancelling the measurement-induced phase. The constant shift $\phi_{\lambda_0}$ in $b(\lambda)$ contributes only an irrelevant global phase. For $\boldsymbol{\mathrm M}=I$, this recovers the familiar technique for removing a CNOT copy of a register by measuring the copy in the $X$ basis and applying $Z^k$ on the main register~\cite{nielsen2010quantum,Gottesman1997Stabilizer}. More generally, without assuming
$S_\Lambda=\operatorname{Im}\boldsymbol{\mathrm M}$, if the populated branch differences span only a subspace
$
D:=
\Span
\{\phi_\lambda\oplus\phi_{\lambda_0}:\lambda\in\Lambda\}
$, we can choose a matrix
$\boldsymbol{\mathrm B}\in\F_2^{n\times m}$
whose columns form a basis of $D$, where $m=\dim D$.
Then
$S_\Lambda=\operatorname{Im}
(\boldsymbol{\mathrm M}\boldsymbol{\mathrm B})$, so a valid phase representation is given by $\boldsymbol{\mathrm H}_\Lambda=(\boldsymbol{\mathrm M}\boldsymbol{\mathrm B})^T$.

For the allowed transformation, looking back at \cref{eq:general-phase-representation}, note that when $m=r_\Lambda$, the kernel $\ker\boldsymbol{\mathrm H}_\Lambda^T$ is trivial, so $z(\lambda)$ vanishes:
$b(\lambda)$ is linear in $\lambda\oplus\lambda_0$,
and hence affine in $\lambda$. When $m>r_\Lambda$, the existence of $z(\lambda)$ permits non-linear transformation on $\lambda$. 

\begin{figure*}
\centering
\includegraphics[width=\linewidth]{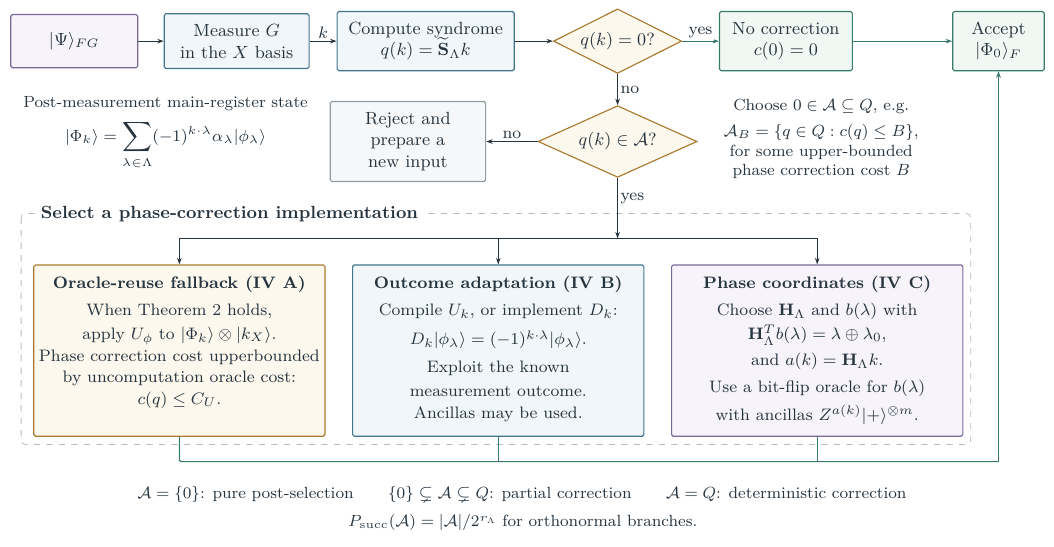}
\caption{\emph{Phase-corrected MBU.} $X$-basis measurement of $G$ yields outcome $k$ and syndrome $q(k)$. The induced phase can be corrected at the implementation cost $c(q)$. Zero syndromes require no correction; accepted nonzero syndromes use oracle reuse under \cref{thm:branch-controlled-deterministic}, outcome-adapted compilation, or optimised phase coordinates. Other outcomes are rejected. Choosing the accepted syndrome set $\mathcal{A}$ interpolates between pure post-selection, partial correction and deterministic MBU. All accepted outputs recover the target state up to a global phase.}
\label{fig:deterministic-mbu-flowchart}
\end{figure*}
\subsection{Trading post-selection against correction cost}
\label{subsec:cost-tradeoff}

The preceding subsections describe how to implement the phase correction
associated with a measurement outcome. We now consider which outcomes
should be accepted when their correction circuits have different costs.
Two outcomes $k$ and $k'$ induce the same relative phase pattern if and
only if $k\oplus k'\in S_\Lambda^\perp$, so the distinct correction tasks
are labelled by $Q=\mathbb F_2^{N_g}/S_\Lambda^\perp$, identified with
$\mathbb F_2^{r_\Lambda}$ through the coordinates $q(k)$ of
\cref{eqn:syndrome_vec}. A protocol chooses an accepted set
$\mathcal A\subseteq Q$: these classes are corrected, whereas all other
runs are rejected. Since the branch states are orthonormal, each class
has probability $2^{-r_\Lambda}$, giving
\begin{equation}
 P_{\mathrm{succ}}(\mathcal A)
 =\frac{|\mathcal A|}{2^{r_\Lambda}},
 \qquad
  N_{\mathrm{rep}}
 =\frac{2^{r_\Lambda}}{|\mathcal A|}.
 \label{eq:generalised-succ-prob}
\end{equation}
The expected number of attempts $N_{\mathrm{rep}}$ assumes independent attempts and the ability to
prepare the input again after rejection. For non-orthogonal branch states, the accepted classes are no longer equiprobable; the exact success probability and its deviation bounds are given in \cref{app:nonorthogonal}. 

Let $c(q)$ be the coherent circuit cost of the selected correction for
syndrome $q$, and $C_U$ the cost of the original uncomputation oracle,
using the same metric, such as gate count or coherent circuit depth.
These are implementation costs, not complexity lower bounds.
Measurement, simple product-state preparation, and classical feed-forward
are accounted for separately. The zero syndrome needs no correction,
so $c(0)=0$, and we always include it in the accepted set $\mathcal A$. Define
\begin{equation}
 \begin{aligned}
 C_{\max}(\mathcal A)&=\max_{q\in\mathcal A}c(q),\\
 \overline C_{\mathrm{corr}}(\mathcal A)
 &=\frac{1}{|\mathcal A|}\sum_{q\in\mathcal A}c(q).
 \end{aligned}
 \label{eq:accepted-correction-costs}
\end{equation}
which are the worst-case and mean accepted correction costs, respectively.
Pure post-selection, $\mathcal A=\{0\}$, has
$C_{\max}(\{0\})=0$ but incurs $N_{\mathrm{rep}} = 2^{r_\Lambda}$ expected attempts.

At the deterministic extreme, $\mathcal A=Q$, all outcomes are accepted.
Crucially, whenever \cref{thm:branch-controlled-deterministic} applies,
the original uncomputation oracle provides a fallback correction for
\emph{every} syndrome. The mechanism is summarised in
\cref{fig:deterministic-mbu-flowchart}: a syndrome-zero outcome requires
no correction, while any other outcome can be corrected by applying
the same oracle to the surviving main register and a garbage register
in $\ket{k_X}$. We use a cheaper outcome-adapted circuit when available
and otherwise fall back to this oracle. Therefore the correction
procedures can always be chosen to satisfy
\begin{equation}
 \begin{aligned}
 c(0)=0,\qquad c(q)\le C_U\quad \forall q\in Q\ \Rightarrow\ 
 C_{\max}(Q) \le C_U.
 \end{aligned}
 \label{eq:deterministic-fallback-bound}
\end{equation}
In particular, no measurement branch needs a coherent correction circuit
more expensive than the original coherent uncomputation. Moreover,
uniformity of the syndrome distribution gives
\begin{equation}
 \overline C_{\mathrm{corr}}(Q)
 =2^{-r_\Lambda}\sum_{q\in Q}c(q)
 \le (1-2^{-r_\Lambda})C_U.
 \label{eq:deterministic-average-bound}
\end{equation}
Thus deterministic MBU never increases the worst-case coherent
correction cost and strictly reduces its expectation. This saving is guaranteed even without finding a simpler correction
for any nonzero syndrome; outcome-dependent compilation can improve it
further.

An intermediate protocol can impose a hard correction budget
$C_{\max}(\mathcal A)\le B$ and maximise the acceptance probability
subject to that constraint. For the specified correction procedures,
the largest such accepted set is
\begin{equation}
 \begin{aligned}
 \mathcal A_B&=\{q\in Q:c(q)\le B\},\\
 C_{\max}(\mathcal A_B)&\le B,\qquad
 P_{\mathrm{succ}}(B)=\frac{|\mathcal A_B|}{2^{r_\Lambda}}.
 \end{aligned}
 \label{eq:correction-budget-acceptance}
\end{equation}
Equivalently, $\mathcal A_B$ maximises $P_{\mathrm{succ}}(\mathcal A)$
over accepted sets satisfying $C_{\max}(\mathcal A)\le B$. When \cref{thm:branch-controlled-deterministic} applies, the budget $B=C_U$ permits deterministic MBU. 
A smaller budget may reject the
syndromes requiring the full uncomputation oracles for phase correction while retaining cheaply correctable
ones, trading shorter worst-case circuits for additional repetitions.
The accepted set need not be a linear subspace, and correction cost need
not follow syndrome Hamming weight. Unlike likelihood-based selection,
all effective syndromes here are equally probable: the selection is
based on their correction costs.

Worst-case circuit cost and total expected work are distinct objectives.
For the latter, take an additive cost model with $C_{\mathrm{pre}}$ for
preparing the incoming state in each attempt and a common per-attempt
measurement and control cost $C_{\mathrm{meas}}$, including any simple
product-state preparation excluded from $c(q)$. Thus the expected output cost is
\begin{equation}
 \begin{aligned}
 C_{\mathrm{out}}(\mathcal A)
 &=\frac{C_{\mathrm{pre}}+C_{\mathrm{meas}}}
         {P_{\mathrm{succ}}(\mathcal A)}
   +\overline C_{\mathrm{corr}}(\mathcal A)\\
 &=\frac{2^{r_\Lambda}(C_{\mathrm{pre}}+C_{\mathrm{meas}})
          +\sum_{q\in\mathcal A}c(q)}{|\mathcal A|}.
 \end{aligned}
 \label{eq:mbu-expected-output-cost}
\end{equation}
The correction term is not multiplied by the repetition overhead,
because rejected attempts execute no correction. With other costs
fixed, adding a rejected class $q$ decreases this expected cost exactly
when $c(q)<C_{\mathrm{out}}(\mathcal A)$; even a relatively expensive
correction can therefore be preferable to restarting the preparation.
For the deterministic MBU under \cref{thm:branch-controlled-deterministic},
\begin{equation}
 C_{\mathrm{out}}(Q)
 \le C_{\mathrm{pre}}+C_{\mathrm{meas}}
      +(1-2^{-r_\Lambda})C_U.
 \label{eq:deterministic-total-bound}
\end{equation}
Compared with the coherent benchmark $C_{\mathrm{pre}}+C_U$, the
fallback guarantee alone yields a net saving whenever
$C_{\mathrm{meas}}<2^{-r_\Lambda}C_U$; cheaper compiled corrections
relax this sufficient condition. Thus the guaranteed reduction in
coherent-circuit cost is distinct from the implementation-dependent
balance of total costs. Support discovery and compilation costs, when
needed, must also be included or amortised over repeated outputs.

\section{Conclusions}
\label{sec:conclusions}
We have developed an algebraic framework for measurement-based uncomputation (MBU) that connects the geometry of the populated garbage support to quantum error correction. By interpreting the measurement outcome of the garbage register as a phase error and the affine support span of the populated garbage strings as the $X$-stabiliser space, we can interpret the relative phases induced by the measurement as the error syndrome. Pure post-selected MBU accepts only zero-syndrome outcomes, so no phase correction is required. For orthonormal main-register branch states, we have shown that the success probability is $2^{-r_\Lambda}$, where $r_\Lambda$ is the affine rank of the populated garbage support. The post-selection overhead is therefore determined by this rank, rather than growing exponentially with the number of garbage qubits.

As an example of this post-selected regime, we applied the framework to the HHL algorithm~\cite{Harrow_2009}, replacing the inverse quantum phase estimation (QPE) block with an $X$-basis measurement of the eigenvalue register, which roughly halves the per-shot circuit depth (see \cref{sec:hhl-postselected-uncomputation}). For structured problems, such as hypercube graph Laplacians, the populated eigenvalues lie on a low-dimensional affine subspace, so the affine rank attains its logarithmic lower bound in the input bandwidth and the total sampling overhead collapses to a manageable, system-size-independent constant factor at fixed input bandwidth and precision.

We also established a structural criterion for deterministic MBU in \cref{thm:branch-controlled-deterministic}. Whenever this condition holds, the original uncomputation oracle provides a fallback correction for every nonzero syndrome, while zero-syndrome outcomes require no correction. Thus the worst-case phase correction cost never exceeds that of the original uncomputation oracle, implying that the average correction achieves a strictly lower cost than the original oracle. 
Adapting the correction to the known measurement outcome can reduce its cost further.
Existing techniques for modular arithmetic~\cite{Luongo2024MBU}, QROM~\cite{Gidney2019Windowed}, and QRAM routing~\cite{Cesa2025FastQRAM} illustrate how the relevant uncomputation blocks fit this criterion and how additional structure can enable simpler correction circuits. More generally, we can accept and correct only a subset $\mathcal A$ of effective syndrome classes based on the acceptable worst-case phase correction cost, giving a total success probability of $|\mathcal A|/2^{r_\Lambda}$. 

The phase-coordinate transformation supplies additional freedom to simplify these circuits. Minimal coordinates identify the independent relative-phase information, whereas redundant representations can be chosen to match the encoding of the main-register data for a cheaper correction oracle. A wider representation can therefore be useful even if it requires additional work qubits. Together with the choice of accepted syndromes, this leads to trade-offs among worst-case correction cost, workspace, and repetitions. These trade-offs allow MBU implementations to be tailored to the particular resource bottlenecks of early fault-tolerant applications. 

A natural direction for future work is to integrate this algebraic framework into quantum compilers. Automatically identifying the affine support span $S_{\Lambda}$ and its orthogonal complement $S_{\Lambda}^\perp$ at compile-time could allow compilers to dynamically replace deep inverse blocks with MBU protocols whenever the rank condition is favourable. Furthermore, compilers could be designed to analyse the uncomputation oracle to detect when it satisfies the conditions of \cref{thm:branch-controlled-deterministic}, enabling the automated generation of deterministic feed-forward corrections. Finally, extending this QEC-based uncomputation perspective to more advanced linear algebra primitives, such as Quantum Singular Value Transformation (QSVT)~\citep{gilyen2019quantum} and block encodings~\citep{chakraborty2019power}, may reveal further structural pathways to compressing algorithm depth for early fault-tolerant machines.

\begin{acknowledgments}
    ZC thanks Zhenhuan Liu and Greg Boyd for inspiring discussions at the early stage of this project.
    PWH acknowledges support from the Engineering and Physical Sciences Research Council (EPSRC) Doctoral Training Partnership (EP/W524311/1) with a CASE Conversion Studentship in collaboration with Quantum Motion. PWH further acknowledges support from the Ministry of Education, Taiwan, for a Government Scholarship to Study Abroad (GSSA) and St. Catherine's College, University of Oxford, for an Alan Tayler Scholarship. ZC acknowledges support from the EPSRC Quantum Technologies Career Acceleration Fellowship (UKRI1226). For the purpose of Open Access, the author has applied a CC BY public copyright license to any Author Accepted Manuscript version arising from this submission. 

    The authors used Claude Opus and Fable, GPT-5.x, and Gemini 2.5/3.1 Pro to assist in drafting and editing portions of the manuscript for language, clarity, and style. Multiple versions of these models were used over the course of manuscript preparation as newer versions became available. All scientific content, interpretation, and conclusions were developed and verified by the authors.
\end{acknowledgments}

\newpage
\appendix

\crefalias{section}{appendix}
\crefalias{subsection}{appendix}
\begin{center}
\noindent{\large\bfseries Appendices}
\end{center}
\appendixtableofcontents

\section{Proof of equivalent statement on the dimension of \texorpdfstring{$S_{\Lambda}$}{S\_Lambda}} \label{sec:proof_equiv_statement}
Here we want to show that $\lambda_0 \not \in S_{\Lambda}$ if and only if there does not exist any odd-sized subset of $\Lambda$ that sums to the zero vector.

First, we will show that if there exists an odd-sized subset of $\Lambda$ that sums to the zero vector, then $\lambda_0 \in S_{\Lambda}$. Let $\Lambda' \subseteq \Lambda$ be the odd-number-size subset of $\Lambda$ that sums to the zero vector, then we have $\bigoplus_{\lambda \in \Lambda'} \lambda = 0$. Note if we sum up the corresponding elements in $S_{\Lambda}$, we have $\bigoplus_{\lambda \in \Lambda'} (\lambda \oplus \lambda_0) = (\bigoplus_{\lambda \in \Lambda'} \lambda) \oplus \lambda_0 = \lambda_0$, where we have used the fact that we are summing up an odd number of $\lambda_0$ and thus we will end up with one $\lambda_0$ left. Hence, we have $\lambda_0 \in S_{\Lambda}$.

Next, we will show that if $\lambda_0 \in S_{\Lambda}$, there must be at least one odd-sized subset of $\Lambda$ that sums to the zero vector. Assume that $\lambda_0 \in S_{\Lambda}$, then we can write $\lambda_0 = \bigoplus_{\lambda \in \Lambda''} (\lambda \oplus \lambda_0)$ for some subset $\Lambda'' \subseteq \Lambda$. Without loss of generality, we can assume $\lambda_0 \notin \Lambda''$, because the term $\lambda_0 \oplus \lambda_0 = 0$ contributes nothing to the sum. This implies that $\bigoplus_{\lambda \in \Lambda''} \lambda = 0$ when the size of $\Lambda''$ is odd and $(\bigoplus_{\lambda \in \Lambda''} \lambda)  \oplus \lambda_0= 0$ when the size of $\Lambda''$ is even. Hence, in both cases, there exists an odd-number-size subset of $\Lambda$, either $\Lambda''$ or $\Lambda''\cup\{\lambda_0\}$, that sums to zero. 

\section{Support discovery and truncation error}
\label{app:sec:support-discovery}

\subsection{Heavy-hitters support discovery}
The post-selection rules developed in \cref{sec:framework} require knowledge of the populated support $\Lambda$ or, more generally, of the subset of garbage strings whose probabilities exceed a prescribed resolution threshold $\mu>0$. Let $p_\lambda$ denote the probability distribution obtained by measuring the garbage register in the computational basis. Support discovery is then the classical heavy-hitters problem of identifying all labels with $p_\lambda\ge\mu$.

Consider a support element with $p_\lambda \ge \mu$. After $N_{\mathrm{samp}}$ independent samples, the probability of never observing this element is
\[
(1-p_\lambda)^{N_{\mathrm{samp}}}
\le
(1-\mu)^{N_{\mathrm{samp}}}
\le
e^{-\mu N_{\mathrm{samp}}},
\]
where the final inequality uses $1-\mu \le e^{-\mu}$. A union bound over the heavy hitters $\Lambda_\mu=\{\lambda:p_\lambda\ge\mu\}$, of which there are at most $1/\mu$, gives
\[
\delta_{\mathrm{fail}}
\le
|\Lambda_\mu|e^{-\mu N_{\mathrm{samp}}}
\le
\frac{1}{\mu}
e^{-\mu N_{\mathrm{samp}}}.
\]
Therefore, to recover every heavy hitter with probability at least $1-\delta_{\mathrm{fail}}$, it suffices to choose
\[
N_{\mathrm{samp}} =\mathcal{O}
\left(
\frac{1}{\mu}
\log\frac{1}{\mu\delta_{\mathrm{fail}}}
\right),
\]
which establishes the sample complexity quoted in \cref{subsubsec:support-discovery}.

In practice, it may be cheaper to sample from a different state that shares the same garbage support but follows a different distribution $q_\lambda$. Suppose the amplification factor of the two distributions is bounded by
\[
\kappa^2=\max_\lambda
\frac{p_\lambda}{q_\lambda}.
\]
Every support element with $p_\lambda \ge \mu$ then satisfies $q_\lambda \ge \mu/\kappa^2$, so support discovery can be performed on the proxy distribution by applying the tightened heavy-hitters threshold $\mu/\kappa^2$. Repeating the argument above~\citep{motwani1995randomized,mitzenmacher2005probability} shows that all such labels can be recovered with high probability using
\begin{equation}
N_{\mathrm{samp}}
=
\mathcal{O}\left(
\frac{\kappa^2}{\mu}
\log\frac{1}{\mu\delta_{\mathrm{fail}}}
\right)
\label{eq:heavy-hitter-samples-proxy}
\end{equation}
samples, without prior knowledge of the distribution. The recovered labels define the estimated support $\hat{\Lambda}$; all components with $p_\lambda < \mu$ are treated as truncation error and contribute the omitted probability mass $\epsilon_{\rm tail}=\sum_{\lambda\notin\hat{\Lambda}}p_\lambda$.

\subsection{Probability mass learning and truncation error}

The heavy-hitters bound guarantees the discovery of the support; bounding the fidelity of the uncomputed state additionally requires an estimate of the omitted probability mass. Because the samples are drawn from the proxy distribution, the empirical frequencies $\hat{q}_\lambda$ are unbiased estimates of the proxy weights $q_\lambda$ rather than of the amplified weights $p_\lambda$. Define the empirical proxy tail mass $\hat{\epsilon}_{\rm init, tail} = 1 - \sum_{\lambda \in \hat{\Lambda}} \hat{q}_\lambda$. Since the amplification from the proxy distribution to the final uncomputed state is bounded by $\kappa^2$, the final tail mass obeys $\epsilon_{\rm tail} \le \kappa^2 \epsilon_{\rm init, tail}$, so the empirical proxy tail provides a worst-case estimate of the final truncation error.

The missed branches penalise the fidelity of the uncomputed state directly: the accepted state differs from the ideal exact-support post-selected state by at most $\mathcal{O}(\sqrt{\epsilon_{\rm tail}})$ in state-vector norm. To target a final amplitude error of $\mathcal{O}(\epsilon)$, it therefore suffices to extend the support-discovery stage until $\epsilon_{\rm tail} = \mathcal{O}(\epsilon^2)$, i.e., $\epsilon_{\rm init, tail} = \mathcal{O}(\epsilon^2/\kappa^2)$.

Once a reliable candidate set $\hat{\Lambda}$ has been obtained, the compiler computes $S_{\Lambda} = \Span_{\F_2}\{\lambda \oplus \lambda_{0}:\lambda\in\hat{\Lambda}\}$ and the acceptance condition $S_{\Lambda}^\perp$ using standard linear algebra over $\F_2$.

\section{Details for QPE uncomputation in HHL}
\label{app:spectral-hhl}

\subsection{Support discovery for HHL}
\label{app:subsec:preliminary_qpe}

The post-selection rule requires prior knowledge of the populated eigenvalue support $\Lambda$; if the estimated support is incomplete, outcomes may be accepted that induce a non-trivial, uncorrected phase on a missing eigenvalue branch, contributing the truncation error analysed in \cref{app:sec:support-discovery}. For HHL, the support is sampled from the state after the forward QPE, $\mathrm{QPE}\ket{b}\approx\sum_{\lambda}\beta_\lambda\ket{\phi_\lambda}\ket{\lambda}$, as illustrated in \cref{fig:hhl-mbu}(b), i.e., from the proxy distribution $q_\lambda=|\beta_\lambda|^2$. Since the controlled rotation amplifies the probabilities by at most the condition number squared, $\kappa^2=(\lambda_{\max}/\lambda_{\min})^2$, the heavy-hitters procedure of \cref{app:sec:support-discovery} applies directly with the tightened threshold $\mu/\kappa^2$ and the sample complexity of \cref{eq:heavy-hitter-samples-proxy}. Once the support estimate $\hat{\Lambda}$ is obtained, the compiler computes $S_\Lambda$ and $S_\Lambda^\perp$ using standard linear algebra over $\F_2$.

\subsection{Work ancilla qubits, tapers, and early fault-tolerant trade-offs}
\label{app:subsec:taper}
\label{sec:EFT_trade_off}

This subsection accounts for the QPE work ancilla qubits and the taper of the eigenvalue register, which were set aside under the perfect-QPE assumption of \cref{sec:hhl-postselected-uncomputation}, and quantifies the resulting space--time trade-off on early fault-tolerant devices, which are constrained primarily by coherent circuit depth and qubit lifetimes.

Since the input state $\ket{b}$ is generally a superposition of eigenvectors, implementations relying on longer-time evolution or higher-precision spectral filtering introduce additional work ancilla qubits beyond the core eigenvalue bits. As noted in \citet{Harrow_2009}, the phase estimation procedure and the subsequent controlled rotation/post-selection on the flag qubit then produces a state of the form
\begin{align*}
    \sum_{\lambda\in\Lambda} c_{\widetilde\lambda|\lambda}\ket{\phi_{\lambda}} \ket{\widetilde\lambda}\ket{\mathrm{aux}(\lambda, \widetilde\lambda)}
\end{align*}
where $c_{\widetilde\lambda|\lambda}$ is concentrated around the true eigenvalue such that $c_{\lambda|\lambda} \approx \alpha_\lambda = 
\widetilde{C}\cdot\beta_\lambda/\lambda$, and the auxiliary state $\lvert\mathrm{aux}(\lambda, \widetilde\lambda)\rangle$ is entangled with the specific eigen-branches. To fully disentangle the workspace during MBU, these work ancilla qubits must therefore be included in the total garbage register. Their populated support in the auxillary register, however, lacks the affine structure of the core eigenvalue readout, so the trivial-syndrome condition $k \in S_{\Lambda}^\perp$ can only be satisfied across the entire superposition by post-selecting these qubits on the all-zero state in the $X$ basis, at a cost that grows exponentially with their number, while accepting valid $k$ strings over the $\ket{\tilde \lambda}$ and rotation success flag registers. The additional cost from sampling the auxillary register is controlled by compressing the auxiliary subspace: preparing the eigenvalue register with a modern windowed taper, such as a Gaussian~\citep{rendon2023lowdepth,chen2025quantum}, Kaiser~\citep{berry2024analyzing}, or discrete prolate spheroidal sequence (DPSS) taper~\citep{patel2026optimal}, rather than the sine taper used in the original HHL paper, bounds the amplitude leakage tightly enough that only $\mathcal{O}(\log_2 \log (1/\epsilon))$ extra work qubits are required for a phase estimation error rate $\epsilon$, keeping the additional sampling cost highly manageable. Any residual amplitude leakage that survives the taper leaves the branch states only approximately orthogonal; the resulting shift of the post-selection success probability away from $2^{-r_\Lambda}$ is controlled by the residual coherence of the garbage state, as quantified in \cref{app:nonorthogonal}.

Replacing the inverse QPE block with purely post-selected MBU then trades circuit depth for repetitions. The per-shot quantum depth is strictly reduced to the forward computation $D_{\rm pre}$, and the $N_g$ phase estimation qubits are released immediately upon measurement. Accounting for the sampling penalty $2^{r_\Lambda}$ of \cref{thm:succ-prob-post-selected-mbu}, with the affine rank $r_\Lambda$ of the discretised eigenvalue support defined in \cref{eqn:affine_rank}, and for the zero-state post-selection of the $\mathcal{O}(\log_2 \log(1/\epsilon))$ taper ancilla qubits, the total expected runtime of the post-selected algorithm scales as
\begin{equation*}
    \mathbb{E}[T_{\rm PS}] = \frac{D_{\rm pre}}{p_{\rm rot}} \cdot 2^{r_\Lambda}\cdot \mathcal{O}\left( \log(1/\epsilon) \right),
\end{equation*}
where $p_{\rm rot}$ is the success probability of the post-selection after the controlled rotation and $\epsilon$ is the target error tolerance. Comparing this to the coherent baseline $\mathbb{E}[T_{\rm coh}] = D_{\rm pre}/p_{\rm rot} + D_{\rm invQPE}$, in which the ancilla is measured immediately after the controlled rotation so that rejected attempts do not execute the inverse QPE, the post-selected HHL formally trades the coherent inverse block for classical repetition. This comparison excludes the one-off support-discovery stage of \cref{app:subsec:preliminary_qpe}, which standard HHL does not require; its sample complexity scales with the $\kappa^2/\mu$ threshold and can therefore be significant for ill-conditioned systems, although this cost is incurred once and amortised over all subsequent runs.

Assuming the controlled-rotation stage in \cref{fig:hhl-mbu} is negligible compared with the forward and inverse QPE blocks, i.e., $D_{\mathrm{rot}} \ll D_{\mathrm{QPE}}$, replacing the inverse QPE with MBU approximately halves the per-shot circuit depth. For the representative hypercube graph example (see \cref{app:subsec:hypercube-theory}) with affine rank $r_\Lambda=3$ and bandwidth $K=4$ (the $n=4$ instance in \cref{fig: r_lambda_vs_n}), post-selection on the eigenvalue register requires an expected $2^{r_\Lambda}=8$ repetitions, and achieving a realistic error tolerance (e.g., $\epsilon = 10^{-3}$) requires strictly post-selecting roughly $5$ additional zero-state taper qubits when using the DPSS taper (see Theorem~2 and Remark~2 of Ref.~\cite{patel2026optimal}), a further factor of $2^{5}=32$. The total number of repetitions is therefore approximately $256\times$ that of the coherent HHL execution, but each repetition executes a circuit of roughly half the depth: against a baseline in which every attempt executes the full coherent circuit including the inverse QPE, the total expected depth is roughly $256\times\tfrac{1}{2}\approx128$ times larger, while against the early-abort baseline $\mathbb{E}[T_{\rm coh}]$ above, in which rejected attempts already skip the inverse QPE, the overhead approaches $256\times$ for $p_{\rm rot}\ll 1$. More crucially, for such highly structured problems, both the affine rank $r_\Lambda$ and the taper overhead $\mathcal{O}(\log_2 \log 1/\epsilon)$ remain independent of the total graph size. As a result, this sampling overhead is a constant factor independent of the problem size, successfully avoiding the exponential overhead typically associated with pure post-selection.

Furthermore, by roughly halving the coherent depth of each shot, MBU halves the probability of a logical error occurring during the run, which allows us to double the tolerable logical error rate and still achieve a similar accuracy. Using the standard logical error rate formula $p_{\mathrm{L}} = A(p/p_{\mathrm{th}})^{\lfloor (d+1)/2 \rfloor}$, where $p$, $p_{\mathrm{th}}$ and $d$ are the physical error rate, error threshold and code distance, respectively, and taking the physical error rate at half the threshold, $p/p_{\mathrm{th}} = 1/2$, allowing a twice larger $p_{\mathrm{L}}$ through MBU enables a reduction of the code distance from $d$ to $d-2$. This can be especially significant in the early fault-tolerant regime with small $d$. For example, for the rotated surface code, a reduction from distance $5$ to $3$ means almost $3$ times fewer physical qubits.

\subsection{Matrix structure, regimes of advantage, and the hypercube spectrum}
\label{app:subsec:hhl-structure}
\label{app:subsec:hypercube-theory}

The sampling cost of post-selected HHL uncomputation is governed by the affine rank $r_\Lambda$ of the populated eigenvalue support. The rank is small precisely when the eigenvalue labels are strongly algebraically correlated, that is, when the spectrum is constrained by a small number of underlying structural parameters, either because it contains only a few distinct values or because the eigenvalues depend on a low-dimensional parametrisation, so that the eigenvalue estimates occupy a low-dimensional subset of $\mathbb{F}_2^{N_{g}}$.

A particularly important unifying class with this property consists of the operators diagonalisable in a Fourier-type basis, including circulant matrices, graph Laplacians, and discretised differential operators. In these cases, the matrix admits a decomposition
\begin{equation}
A = F^\dagger \,\mathrm{diag}(\lambda_{1}, \lambda_2, \dots)\, F,
\end{equation}
where $F$ is a Fourier or Fourier-like transform and the eigenvalues $\lambda_k$ are indexed by a frequency variable $k$ satisfying $\lambda_i \leq \lambda_{j}$ if $i < j$. When $\ket{b}$ is smooth in the computational basis, meaning that its amplitudes vary slowly with the index, its Fourier coefficients decay rapidly with frequency, so the state is well-approximated by a superposition of low-frequency modes,
\begin{equation}
\ket{b} \approx \sum_{k \le K} \beta_k \ket{\phi_{k}},
\quad K \ll n.
\end{equation}

Since $\Lambda\subseteq\lambda_0\oplus S_\Lambda$, the affine rank always obeys $r_\Lambda\ge\log_2|\Lambda|$, so for $K$ populated modes the smallest possible rank is logarithmic in $K$. When the labels of the populated modes form consecutive integers, the affine rank attains its logarithmic lower bound,
\begin{equation}
r_\Lambda = \Omega (\log_2 K),
\end{equation}
typically much smaller than $N_{g}$. For discretised spectra without such algebraic structure, the rank can instead approach $\min(K, N_{g})$ even for smooth inputs. Because this framework relies on the concentration of $\ket{b}$ in the eigenbasis, the algorithm operates effectively as a semi-dequantised version of HHL; the extent to which this assumption resists full dequantisation is discussed in \cref{app:subsec:dequantization}.

The hypercube graph $Q_n$, the primary benchmark for our numerical validations, instantiates this structure exactly. An $n$-dimensional hypercube consists of $2^n$ vertices, each represented by a binary string of length $n$, with two vertices connected if their Hamming distance is exactly one. Its Laplacian is a sum of $n$ commuting operators, each acting on a single dimension; rescaled so that the eigenvalues are integers, it reads
\begin{equation}A = \sum_{i=1}^n \left(I \otimes \dots \otimes \left[\frac{I - X}{2}\right]_i \otimes \dots \otimes I\right).\end{equation}
Because the eigenvalues of each individual term $\frac{1}{2}(I-X)$ are $\{0, 1\}$, the eigenvalues of $A$ are determined strictly by the Hamming weight $k$ of the corresponding frequency index,
\begin{equation}\lambda_k = k, \quad k \in \{0, 1, \dots, n\},\end{equation}
with multiplicity $\binom{n}{k}$. 

As is standard for HHL, the linear system is solved on the complement of the kernel: the populated support $\Lambda$ is taken over the \emph{nonzero} eigenvalues, the kernel component of $\ket{b}$ is removed by the post-selection of the controlled-rotation ancilla, whose rotation angle vanishes on the zero label, and the condition number $\kappa$ is taken over the populated nonzero eigenvalues. Support discovery, which samples before the controlled rotation, may nevertheless include the zero label in the estimated support. This is harmless, since enlarging the estimated support only enlarges $S_\Lambda$ and makes the acceptance rule stricter, changing the affine rank by at most one; for the hypercube labels, $\{0,1,\dots,K\}$ and $\{1,\dots,K\}$ share the $\lfloor\log_2 K\rfloor+1$ scaling, and our numerics include the zero label accordingly.

\begin{figure}
\centering
\includegraphics[width=\linewidth]{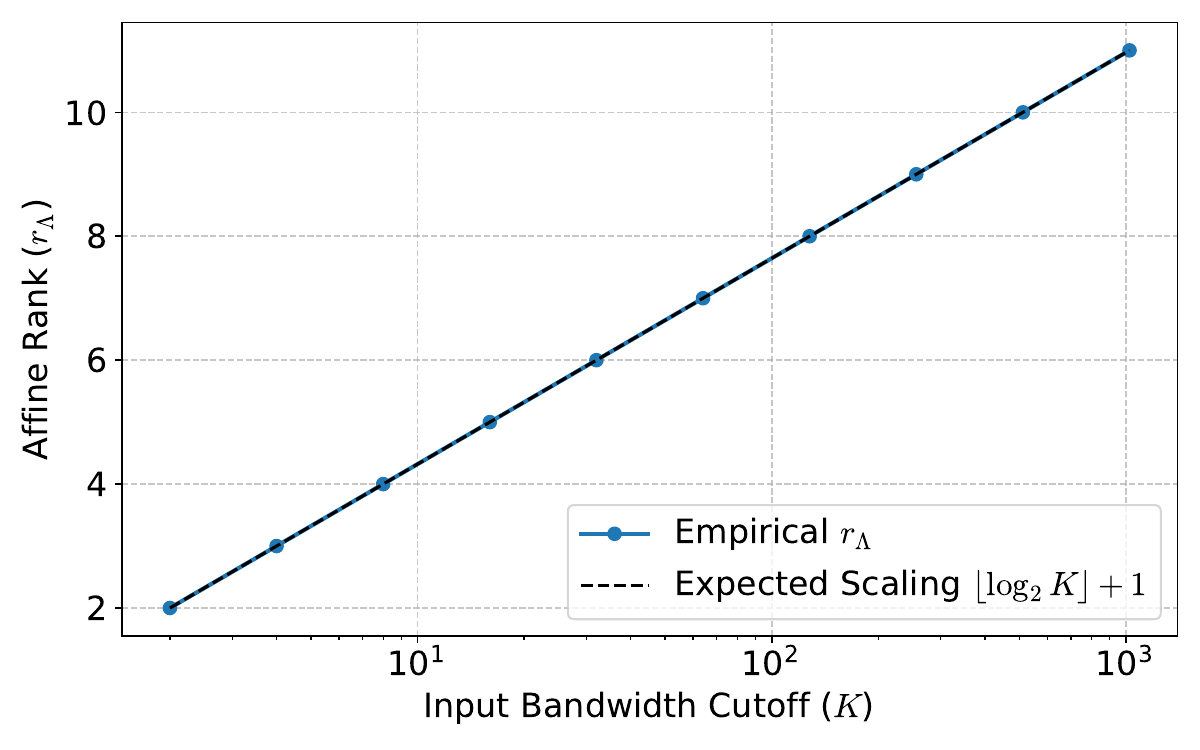}
\caption{\emph{Affine rank $r_{\Lambda}$ versus input bandwidth $K$ for the hypercube graph.} The empirical rank closely follows the expected $\lfloor \log_2 K \rfloor + 1$ scaling, confirming that the dimension of the affine support span is determined solely by the bit-length required to
encode the active frequency modes.}
\label{fig: r_lambda_vs_K}
\end{figure}

For a smooth input with bandwidth $K$, the populated support $\Lambda$ consists of the binary representations of the integers $\{0, 1, \dots, K\}$, discrete samples of the linear function $\lambda_k = k$. While the number of vertices grows exponentially with $n$, the number of active eigenvalues grows only linearly with the bandwidth $K$, so the affine rank scales as $r_\Lambda = \lfloor \log_2 K \rfloor + 1$, the bit-length required to encode the active frequency modes, independently of the system size. This scaling is confirmed numerically in \cref{fig: r_lambda_vs_K} and provides the mathematical foundation for the system-size-independent sampling overhead observed in \cref{sec:numerics}. The exponential input family of \cref{sec:numerics} makes this bandwidth quantitative. Because the graph distance from the source is the Hamming weight, the input $b_x\propto e^{-\gamma d(x,x_0)}$ is a product state, and its weight on the Hamming-weight-$w$ eigenspace is binomial, $\Pr(w)=\binom{n}{w}p^{w}(1-p)^{n-w}$ with $p=(1-e^{-\gamma})^{2}/[2(1+e^{-2\gamma})]$. At fixed $\gamma$ the mean populated weight $np$ grows linearly with $n$, so a fixed retained bandwidth eventually captures a shrinking fraction of the input weight; the bandwidth stays bounded as $n$ grows only under the smoothness scaling $\gamma=\mathcal{O}(n^{-1/2})$, corresponding to signals whose correlation length grows with the graph. Over the tested range $n=4$--$10$ at $\gamma=1.2$, the omitted probability mass beyond the retained support, computable in closed form for this family, remains at or below the per-cent level ($\approx 1\%$ at $n=10$). Even at fixed $\gamma$, the bandwidth grows at most linearly in $n$, so the rank floor $\lfloor\log_2 K\rfloor+1$ grows only logarithmically and the sampling overhead $2^{r_\Lambda}$ remains polynomial in the number of qubits (polylogarithmic in the graph size $N$). Note also that the bandwidth counts \emph{distinct} eigenvalues: each populated weight-$w$ eigenspace is $\binom{n}{w}$-fold degenerate, but only the number of distinct eigenvalue labels enters the affine rank.

\subsection{Dequantisation considerations}
\label{app:subsec:dequantization}

A natural question is whether the assumptions underlying the post-selected HHL algorithm admit classical dequantisation. Recent work, pioneered by Tang~\cite{Tang_2019}, has shown that many quantum machine learning algorithms can be simulated efficiently classically if the input state admits efficient classical $\ell_2$-norm sampling access. In these dequantisation frameworks, the classical algorithm leverages a specialised data structure that allows it to query entries and sample from the distribution $\mathcal{D}_x(i) = |x_i|^2/\|x\|^2$ in time polylogarithmic in the dimension, perfectly mirroring the capabilities of quantum state preparation.

At first glance, the present framework appears to rely on a form of sparsity, namely the concentration of $|b\rangle$ in the eigenbasis of $A$. However, this notion of sparsity is fundamentally different from the assumptions used in canonical dequantisation results. Existing dequantisation techniques typically require the efficient $\ell_2$-norm sampling data structure to exist in the computational basis.

In our setting, sparsity in a Fourier-type eigenbasis does not imply sparsity in the computational basis. In general, if
\begin{equation}
\ket{b} = \sum_{k} \beta_{k} \ket{\phi_{k}}, \quad \ket{\phi_{k}} = F^\dagger \ket{k}
\end{equation}
then a state that is sparse in the $\{\ket{\phi_{k}}\}$ basis is typically highly delocalised and dense in the computational basis $\{\ket{k}\}$.

Therefore, to fully dequantise the post-selected HHL algorithm for these structured operators, one would have to assume the existence of an efficient $\ell_2$-norm classical sampler directly in the Fourier basis, or have further restrictions such that $A$ also admits some level of sparsity in the computational basis~\citep{huang2026classical}. This is a more non-trivial assumption than standard computational-basis sampling. While the quantum computer naturally accesses this basis via the Quantum Fourier Transform during phase estimation, it is not known how to construct, in general, a classical data structure capable of sublinear $\ell_2$-norm sampling of the Fourier coefficients of a generic smooth signal.

We therefore view the present framework as occupying an intermediate regime between fully quantum and fully dequantised linear solvers. The algorithm exploits spectral structure that is naturally accessible in the quantum setting, while remaining robust against classical simulation based on input sparsity.

\subsection{Failure of the phase-correction condition for the inverse QPE}
\label{app:hhl-deterministic-failure}

The inverse QPE illustrates an uncomputation oracle that has the block-diagonal form of \cref{eq:block-form} but does not necessarily satisfy the branch-independent form of \cref{eq:block-form-independent_gen}. Applying the inverse QPE after the garbage measurement therefore cancels the measurement-induced phases, but generally fails to disentangle the garbage register.

Consider a perfect inverse phase estimation oracle for uncomputation,
\begin{equation}
    U_\phi \ket{\phi_\lambda}\ket{\lambda} = \ket{\phi_\lambda}\ket{0}.
\end{equation}
Since this only specifies the action on the populated inputs, the analysis of \cref{sec:oracle condition} applies: on the branch subspace the oracle decomposes as
\begin{equation}
    U_\phi \Pi_\Lambda = \sum_{\lambda
    \in \Lambda }\ketbra{\phi_\lambda}{\phi_\lambda} \otimes W_\lambda X^\lambda,
\end{equation}
where each residual $W_\lambda$ is a unitary satisfying $W_\lambda\ket{0}=\ket{0}$, i.e., $W_\lambda = \ketbra{0}{0} \oplus T_\lambda$ for some unitary $T_\lambda$ acting on the orthogonal complement of $\ket{0}$. For the same oracle to correct the phases of every measurement outcome, \cref{thm:branch-controlled-deterministic} requires $T_\lambda = T$ for all $\lambda$. The additional freedom $U_{s,F}$ in \cref{eq:block-form-independent_gen} does not relax this requirement: the inverse QPE preserves each populated branch, and, as shown in \cref{sec:oracle condition}, $U_{s,F}$ then necessarily acts as the identity on the populated branch subspace. Whether the common-residual condition holds depends on the actual implementation of the phase estimation circuit; we now show that it fails for the textbook implementation built from controlled evolutions and the quantum Fourier transform.

Assume the populated eigenphases are exact Fourier nodes, $\lambda \in \{0,\dots,2^{N_g}-1\}$, so that textbook phase estimation is perfect in this instance. Before the inverse quantum Fourier transform in the forward QPE, the computational-basis state $\ket{j}$ of the eigenvalue register carries the phase $e^{2\pi i j\lambda/2^{N_g}}$. Running QPE backwards therefore gives
\begin{equation}
U_{\phi}\left(\ket{\phi_{\lambda}}\otimes\ket{\lambda}\right)
=
\ket{\phi_{\lambda}}\otimes V_{\lambda\lambda}\ket{\lambda},
\quad
V_{\lambda\lambda}
=
H^{\otimes N_g}\,P_{-\lambda}\,\mathrm{QFT},
\label{eq:app-hhl-inverse-qpe}
\end{equation}
where $\mathrm{QFT}$ and the Hadamard layer act on the garbage register and the diagonal phase operator $P_{-\lambda}\ket{j}=e^{-2\pi i j\lambda/2^{N_g}}\ket{j}$ undoes the phases of the branch $\lambda$ from the controlled evolutions. Since $V_{\lambda\lambda}$ acts only on the eigenvalue register, the inverse QPE does not mix different eigen-branches, so $U_\phi$ is block-diagonal with respect to the populated states $\{\lvert\phi_\lambda\rangle\}_{\lambda\in\Lambda}$, as in \cref{eq:block-form}.

It remains to examine the residual $W_\lambda$ after factoring out the bitwise operation $X^\lambda$. Using the identity $P_{-\lambda}\,\mathrm{QFT}=\mathrm{QFT}\,\mathrm{Sh}_{-\lambda}$, where $\mathrm{Sh}_{c}\ket{\mu}=\ket{\mu+c \bmod 2^{N_g}}$ is the modular shift, the residual evaluates to
\begin{equation}
W_{\lambda}
= V_{\lambda\lambda} X^\lambda =
H^{\otimes N_g}\,\mathrm{QFT}\,\mathrm{Sh}_{-\lambda}X^{\lambda},
\label{eq:app-hhl-residual}
\end{equation}
which satisfies $W_\lambda\ket{0} = \ket{0}$ as required. The dependence of $W_\lambda$ on $\lambda$ is contained entirely in ${\rm Sh}_{-\lambda}X^\lambda$. These two operations implement different additions of the eigenvalue label: the Pauli operation $X^\lambda$ performs bitwise addition, $X^\lambda\ket{\mu} = \ket{\mu\oplus\lambda}$, whereas ${\rm Sh}_{-\lambda}$ performs modular integer subtraction, so their composition acts as
$${\rm Sh}_{-\lambda}X^\lambda\ket{\mu} = \ket{(\mu\oplus\lambda)-\lambda \pmod{2^{N_g}}}.$$
Bitwise XOR and modular addition are not equivalent in general because the latter contains carry operations. Consequently, $W_\lambda$, and hence $T_\lambda$, depends non-trivially on the eigenvalue label $\lambda$.

The inverse QPE therefore cancels the measurement-induced relative phases, but the resulting state generally remains entangled between the main and garbage registers through the branch-dependent residual states $W_\lambda\ket{k_X}$. While we cannot rule out a phase estimation circuit whose residual is branch-independent, conventional implementations of phase estimation cannot be reused for deterministic phase correction. This is why the HHL example of \cref{sec:hhl-postselected-uncomputation} uses post-selected rather than deterministic MBU.

\section{Conditions on the uncomputation oracle}
\label{sec:oracle condition}

We examine the additional assumptions under which a branch-preserving
uncomputation oracle has the structure used in
\cref{thm:branch-controlled-deterministic}.
Suppose the oracle clears every populated branch:
\begin{equation}
 U_\phi\ket{\phi_\lambda}\ket{\lambda}
 =\ket{\phi_\lambda}\ket{0},
 \qquad \lambda\in\Lambda.
 \label{eq:uncomp-unitary}
\end{equation}
Preservation of inner products gives
\begin{equation}
 \bra{0} \braket{\phi_\mu}{\phi_\lambda} \ket{0}= \braket{\phi_\mu}{\phi_\lambda}
 =\braket{\phi_\mu}{\phi_\lambda}\braket{\mu}{\lambda} = \delta_{\mu\lambda},
 \label{eq:app-orthonormality}
\end{equation}
so the normalised branch states must be mutually orthogonal.

Let $\Pi_\Lambda=\sum_{\lambda\in\Lambda}\ketbra{\phi_\lambda}$
act on $F$, and extend the branch states to an orthonormal basis
$\{\ket{\phi_\mu}\}$ of $\mathcal H_F$.
The oracle can be decomposed as
\begin{equation}
 U_\phi(\Pi_\Lambda\otimes I_G)
 =\sum_{\lambda\in\Lambda,\,\mu}
 \ketbra{\phi_\mu}{\phi_\lambda}\otimes V_{\mu\lambda},
 \label{eq:app-block-decomposition}
\end{equation}
where $\mu$ runs over the full basis and
$V_{\mu\lambda}=(\bra{\phi_\mu}\otimes I_G)U_\phi
(\ket{\phi_\lambda}\otimes I_G)$ acts on $G$.
The branchwise erasure condition implies
\begin{equation}
 V_{\mu\lambda}\ket{\lambda}
 =\delta_{\mu\lambda}\ket{0}.
 \label{eq:app-block-constraints}
\end{equation}
Now assume that every block $V_{\mu\lambda}$ is either unitary or zero.
Since a unitary cannot annihilate a nonzero vector, all off-diagonal
blocks vanish. Writing $W_\lambda=V_{\lambda\lambda}X^\lambda$ gives
\begin{equation}
 U_\phi(\Pi_\Lambda\otimes I_G)
 =\sum_{\lambda\in\Lambda}\ketbra{\phi_\lambda}
   \otimes W_\lambda X^\lambda,
 \qquad W_\lambda\ket{0}=\ket{0},
 \label{eq:block-form}
\end{equation}
with each $W_\lambda$ unitary. This block structure uses an additional
assumption: branchwise erasure alone constrains the oracle only on the
correlated inputs $\ket{\phi_\lambda}\ket{\lambda}$, not on the full
space $\operatorname{ran}\Pi_\Lambda\otimes\mathcal H_G$.

For an oracle in this restricted family, its action after the prescribed
$X$-basis measurement is
\begin{equation}
 U_\phi\bigl(\ket{\Phi_k}\otimes\ket{k_X}\bigr)
 =\sum_{\lambda\in\Lambda}\alpha_\lambda
   \ket{\phi_\lambda}\otimes W_\lambda\ket{k_X},
 \label{eq:auto-phase-cancellation}
\end{equation}
using $X^\lambda\ket{k_X}=(-1)^{k\cdot\lambda}\ket{k_X}$.
The explicit measurement-induced signs cancel, but the residual garbage
can still distinguish the branches. Since the branch states are
orthogonal and every populated amplitude is nonzero, recovery of
$\ket{\Phi_0}$ up to a global phase requires
$W_\lambda\ket{k_X}=\ket{w_k}$ independent of $\lambda$.
If this is required for every outcome $k$, equality on the complete
$X$ basis implies $W_\lambda=W$ for all $\lambda$.
Thus, within the family \cref{eq:block-form}, a common residual is both
necessary and sufficient for applying the same oracle after every
measurement outcome to recover the target.

The general form \cref{eq:block-form-independent_gen} extends this family: composing \cref{eq:block-form} with a common residual $W_\lambda=W$ by any unitary $U_{s,F}$ that stabilises the target state, $U_{s,F}\ket{\Phi_0}=\ket{\Phi_0}$, preserves the phase-correction property, but the composite oracle is generally no longer branch-preserving, since $U_{s,F}$ only has to fix the superposition $\ket{\Phi_0}$ rather than each branch state. Conversely, if the oracle is required to preserve each populated branch, then $U_{s,F}\ket{\phi_\lambda}=e^{i\theta_\lambda}\ket{\phi_\lambda}$, and stabilising $\ket{\Phi_0}=\sum_{\lambda\in\Lambda}\alpha_\lambda\ket{\phi_\lambda}$ with every $\alpha_\lambda$ nonzero forces $e^{i\theta_\lambda}=1$ for all $\lambda\in\Lambda$. Hence $U_{s,F}$ acts as the identity on $\Pi_\Lambda$, and within the branch-preserving family \cref{eq:block-form} the common residual remains the full characterisation.

\section{Deterministic measurement-based uncomputation examples}
\label{app:deterministic-examples}

This appendix shows how the deterministic examples of \cref{sec:correctable_framework} map onto our framework and where each construction instantiates \cref{thm:branch-controlled-deterministic}; the full constructions are given in the original references. In all three cases every effective syndrome class is accepted, $\mathcal{A}=Q$ with $P_{\mathrm{succ}}=1$.

\subsection{Modular arithmetic: Comparator qubits in modular additions}
\label{app:deterministic-modular}

In the modular-addition circuits of Ref.~\cite{Luongo2024MBU}, the garbage is a single comparator qubit storing the Boolean predicate $g(x,y)=\mathbbm{1}[\,x+y\ge p\,]$ that decides the conditional subtraction, with the modular sum $z(x,y)=x+y-g(x,y)p$ held in the surviving arithmetic registers. In the notation of \cref{sec:framework}, the populated garbage support is $\Lambda=\{0,1\}$ with affine rank $r_\Lambda=1$, so post-selecting the syndrome-zero outcome succeeds with probability $1/2$ by \cref{thm:succ-prob-post-selected-mbu}, reproducing the success probability of the single-qubit MBU lemma of Ref.~\cite{Luongo2024MBU}. The comparator oracle is the branch-controlled bit flip $\sum_{x,y}\ketbra{x,z(x,y)}\otimes X^{g(x,y)}$, i.e., \cref{eq:block-form} with $W_\lambda=I$, and the correction of the non-trivial syndrome $k=1$ is exactly the diagonal $D_k=\sum_{x,y}(-1)^{k\,g(x,y)}\ketbra{x,z(x,y)}$ of \cref{eq:main-register-diagonal}, compiled as phase kickback of one additional evaluation of $U_g$ on a $\ket{-}$ ancilla together with two Hadamard gates and one NOT gate, avoiding a coherent reversal of the comparator circuit. In the circuit of Ref.~\cite{Luongo2024MBU} this ancilla is simply the recycled measured qubit itself, re-prepared by an outcome-controlled Hadamard, while the trailing Hadamard and NOT reset the branch-independent residual to $\ket{0}$; discarding the measured qubit and preparing a fresh $\ket{-}$ ancilla instead would implement the identical channel.

\subsection{Windowed lookup: Unary encoding and fixup tables}
\label{app:deterministic-lookup}

In the QROM unlookup of Ref.~\cite{Gidney2019Windowed}, the garbage register holds the $N_g$-bit table value $T(a)$ for each address $a=(u,v)$ in the surviving address register, with the address split into two non-overlapping blocks satisfying $2^{|u|}\approx 2^{|v|}\approx\sqrt{L}$ for a table of size $L$. Measuring the garbage register in the $X$ basis with outcome $k$ calls for the diagonal correction $D_k\ket{u,v}=(-1)^{k\cdot T(u,v)}\ket{u,v}$ of \cref{eq:main-register-diagonal}. Because the lookup table is available at compile time and $k$ is classically known at correction time, this phase can be recomputed from the surviving address register alone through the reduced fixup tables
\begin{equation}
F_{k,v}(u)
=
k\cdot T(u,v)
\pmod 2.
\label{eq:app-lookup-fixup-table}
\end{equation}
Converting the $v$-block into a unary register, $\ket{v}\ket{0^{2^{|v|}}}\mapsto\ket{v}\ket{e_v}$, lets the unary register select which fixup subtable $F_{k,v}$ is active while $u$ indexes within it; after the signs are applied, the unary register is cleared. The resulting cost $O(2^{|u|}+2^{|v|})$ is minimised by the equal split $|u|\approx|v|\approx\frac{1}{2}\log_2 L$ to $O(\sqrt{L})$, rather than the $O(L)$ of coherently reversing the full lookup. In terms of \cref{subsec:deterministic-criterion}, the unlookup oracle is $U_\phi=\sum_a\ketbra{a}\otimes X^{T(a)}$, i.e., \cref{eq:block-form} with $W_\lambda=I$. Each $D_k$ is thus compiled as a cheaper oracle than $U_\phi$ itself, with the unary register serving as internal working space of this compilation; the measured garbage register is discarded immediately, with no further operation applied to it.

\subsection{QRAM routing: Adaptive Clifford inversion of \texorpdfstring{$U_{\mathrm{NOHE}}$}{U\_NOHE}}
\label{app:deterministic-qram}

\citet{Cesa2025FastQRAM} compiles bucket-brigade-type QRAM~\cite{BB-QRAM} by expanding the address register into a one-hot pointer register together with an auxiliary nested one-hot encoding (NOHE) register, generated by an isometry
$U_{\mathrm{NOHE}}^{(N)}:
\mathcal H^{(\log N)}
\rightarrow
\mathcal H^{(N-1)}$. After the memory query, the NOHE register plays the role of garbage that must be removed, and a direct implementation of the inverse would require coherently reversing the non-Clifford routing gadgets used to construct $U_{\mathrm{NOHE}}^{(N)}$. Instead, Ref.~\cite{Cesa2025FastQRAM} replaces this coherent inverse by adaptive single-qubit Pauli measurements with Pauli feed-forward, exploiting the fact that only a right inverse on the image of the encoding is required. Every measurement branch is accepted, so the protocol realises a deterministic MBU of the NOHE register. For the minimal non-trivial example $N=4$, with two address qubits $(i,j)\in\F_2^2$, the encoding is
\begin{equation}
U_{\mathrm{NOHE}}^{(4)}
=
\sum_{i,j\in\F_2}
\ket{i,\;j\oplus ij,\;ij}\bra{i,j},
\label{eq:app-qram-UNOHE4}
\end{equation}
and the right inverse used in Ref.~\cite{Cesa2025FastQRAM} is $G=\sum_{i,j,k\in\F_2}\ket{i,\;j\oplus k}\bra{i,j,k}$, which satisfies $G\,U_{\mathrm{NOHE}}^{(4)}\ket{i,j}=\ket{i,j}$ on the image but is not a unitary operation on the full three-qubit space. We now show that the measurement-based implementation of $G$ is a direct instance of \cref{thm:branch-controlled-deterministic}.

\subsubsection{Mapping onto \texorpdfstring{\cref{thm:branch-controlled-deterministic}}{Theorem 2} via dilation}
\label{app:subsec:qram-mapping}

The $N=4$ example can be placed entirely within the structural criterion of \cref{subsec:deterministic-criterion} by making the dilation of the encoding isometry explicit. Any physical implementation of $U^{(4)}_{\mathrm{NOHE}}$ acts unitarily on the two address qubits together with a fresh ancilla; one realisation is
\begin{equation}
\tilde U = \mathrm{CNOT}_{3\to 2}\,\mathrm{Toffoli}_{12\to 3},
\qquad
\tilde U\ket{i,j,0} = \ket{i,\ j\oplus ij,\ ij},
\label{eq:app-qram-dilated}
\end{equation}
so the isometry is the restriction of $\tilde U$ to the ancilla-in-$\ket{0}$ slice, and resetting the third qubit to $\ket{0}$ is operationally identical to implementing the right inverse $G$ on the image. The dilated uncomputation unitary factors as
\begin{equation}
\tilde U^\dagger
=
\underbrace{\mathrm{Toffoli}_{12\to 3}}_{\text{branch-controlled bit flip}}
\;
\underbrace{\mathrm{CNOT}_{3\to 2}}_{\text{branch-independent Clifford}} .
\label{eq:app-qram-factorization}
\end{equation}
The Clifford part is branch-independent and cheap, so it is applied coherently; its role is to restore the diagonal-garbage form of \cref{eq:framework-state}: on the image state $\ket{\chi}=\sum_{i,j}\alpha_{ij}\ket{i,\ j\oplus ij,\ ij}$ it yields
\begin{equation}
\mathrm{CNOT}_{3\to2}\ket{\chi}
=
\sum_{i,j}\alpha_{ij}\ket{i,j}\otimes\ket{i\wedge j},
\label{eq:app-qram-unmixed}
\end{equation}
with the branch label $\lambda=g(i,j)=i\wedge j$, a Boolean function of the surviving branch states. The remaining factor is the branch-controlled bit-flip oracle
\begin{equation}
\mathrm{Toffoli}_{12\to3}
=
\sum_{i,j}\ketbra{i,j}\otimes X^{\,i\wedge j},
\label{eq:app-qram-toffoli-block}
\end{equation}
i.e., \cref{eq:block-form} with $W_\lambda=I$. \cref{thm:branch-controlled-deterministic} therefore directly applies: measuring the third qubit in the $X$ basis with outcome $s_0\in\F_2$ and applying the correction oracle to the collapsed garbage induces the diagonal phase $(-1)^{s_0\,(i\wedge j)}$, i.e., a $\mathrm{CZ}^{s_0}$ on the two surviving qubits, after which the state equals $\sum_{i,j}\alpha_{ij}\ket{i,j}=G\ket{\chi}$ deterministically for both outcomes. This is nothing but the single-qubit MBU mechanism of \cref{app:deterministic-modular} with the predicate $g=\mathrm{AND}$: the expensive non-Clifford Toffoli is exactly what the measurement eliminates.

Three features of the protocol of Ref.~\cite{Cesa2025FastQRAM} become transparent here. First, the promise that the input state lies in the image of the encoding is the framework's restriction to the branch subspace: \cref{thm:branch-controlled-deterministic} constrains the correction oracle only on $\Pi_\Lambda\otimes\cH_G$, so the behaviour of the protocol off the image is irrelevant; this is why a right inverse on the image suffices. Second, the cancellation, on the image, of the anomalous phase appearing in the branch analysis of Ref.~\cite{Cesa2025FastQRAM} is the same syndrome bookkeeping that cancels the explicit signs in \cref{eq:auto-phase-cancellation}. Third, the adaptive single-qubit measurements with Pauli feed-forward constitute a hardware-level compilation of the measure-then-correct step, whose mechanism we now make explicit.

The key to this compilation is the role of the two ancilla qubits in Fig.~6(b) of Ref.~\cite{Cesa2025FastQRAM}. They are logically distinct from the dilation: the dilation ancilla is the third NOHE qubit, which the encoding has already created physically, so no qubit needs to be added to make the isometry unitary. The ancilla pair is instead a gate-teleportation gadget whose purpose is to apply the feed-forward correction $D_{s_0}=\mathrm{CZ}^{s_0}$ without any adaptively applied entangling gate. Denote the surviving qubits by $q_1,q_2$ and the ancilla pair by $a_1,a_2$, and let $\ket{\psi}$ be the state held by $(q_1,q_2)$ when the ancilla qubits are measured, i.e., after the CNOT and the $X$-basis measurement of the third qubit with outcome $s_0$. The ancilla qubits are prepared in $\ket{+}\ket{+}$ and entangled by the outcome-independent Clifford gates $\mathrm{CZ}_{a_1a_2}$, $\mathrm{CZ}_{q_1a_1}$, and $\mathrm{CZ}_{q_2a_2}$, producing the cluster-like state
\begin{equation}
\ket{\Xi}
=
\frac{1}{2}\sum_{a,b\in\F_2}(-1)^{ab}\left(Z^{a}\otimes Z^{b}\ket{\psi}\right)_{q_1q_2}\otimes\ket{a,b}_{a_1a_2}.
\label{eq:app-qram-cluster}
\end{equation}
Measuring both ancilla qubits in the $Z$ basis with outcomes $s_1=(\alpha,\beta)\in\F_2^2$ selects the single term $(a,b)=(\alpha,\beta)$,
\begin{equation}
\bra{\alpha,\beta}_{a_1a_2}\ket{\Xi}\propto \left(Z^{\alpha}\otimes Z^{\beta}\right)\ket{\psi},
\label{eq:app-qram-gadget-Z}
\end{equation}
so the gadget acts as the identity up to the Pauli byproduct $Z^{\alpha}\otimes Z^{\beta}$. Measuring both ancilla qubits in the $X$ basis instead yields
\begin{align}
\bra{\alpha_X,\beta_X}_{a_1a_2}\ket{\Xi}
&\propto
\sum_{a,b\in\F_2}(-1)^{ab+\alpha a+\beta b}\left(Z^{a}\otimes Z^{b}\right)\ket{\psi} \nonumber \\
&\propto
\mathrm{CZ}_{q_1q_2}\left(Z^{\beta}\otimes Z^{\alpha}\right)\ket{\psi},
\label{eq:app-qram-gadget-X}
\end{align}
where the second step follows by evaluating the coefficient of each computational basis state $\ket{x,y}$ of $(q_1,q_2)$: the double sum equals $2(-1)^{(x\oplus\alpha)(y\oplus\beta)}$, and $(x\oplus\alpha)(y\oplus\beta)=xy\oplus\beta x\oplus\alpha y\oplus\alpha\beta$. The gadget therefore teleports the pre-deposited $\mathrm{CZ}$ onto the surviving qubits, up to the Pauli byproduct $Z^{\beta}\otimes Z^{\alpha}$. Choosing the ancilla measurement basis conditioned on $s_0$, namely the $Z$ basis for $s_0=0$ and the $X$ basis for $s_0=1$, is thus precisely the feed-forward application of the correction $D_{s_0}=\mathrm{CZ}^{s_0}$ of \cref{eq:main-register-diagonal}, and the byproducts reproduce the Pauli corrections of Ref.~\cite{Cesa2025FastQRAM},
\begin{equation}
C_{0,(\alpha,\beta)}
=
Z^\alpha\otimes Z^\beta,
\qquad
C_{1,(\alpha,\beta)}
=
Z^\beta\otimes Z^\alpha,
\label{eq:app-qram-corrections}
\end{equation}
including the swap $\alpha\leftrightarrow\beta$ between the two branches. All entangling operations are outcome-independent and can be applied before any measurement; the only adaptive operations in the entire uncomputation are single-qubit measurement-basis choices and the final Pauli-frame update, which is what makes this compilation attractive at the hardware level.

For general $N$, the nested one-hot encoding is built recursively from levels of the same form, i.e., Toffoli-type indicator computations onto fresh ancilla qubits followed by branch-independent CNOT layers that restore the diagonal-garbage form, so the factorisation of \cref{eq:app-qram-factorization} applies level by level, with all corrections remaining diagonal phases computable from the data held in the surviving pointer register. We note that this mapping uses a mild extension of the pipeline of \cref{subsec:deterministic-criterion}: a fixed (branch-independent) Clifford part is applied coherently before the $X$-basis measurement. Since this Clifford part merely redefines the pre-uncomputation state, the criterion applies unchanged to the transformed state.

\section{Non-orthogonal branch states and robustness}
\label{app:nonorthogonal}

The analysis in the main text focuses on the setting, where the branch states $\{\ket{\phi_{\lambda}}\}_{\lambda\in\Lambda}$ are orthogonal to each other. Under this condition, each $X$-basis measurement outcome occurs with equal probability, and the success probability is governed by the fraction of accepted syndrome classes, $P_{\mathrm{succ}}(\mathcal{A})=|\mathcal{A}|/2^{r_{\Lambda}}$ (\cref{eq:generalised-succ-prob}). We now extend the framework to the general case, where the branch states need not be mutually orthogonal. In this setting, the reduced garbage state
\begin{equation}
\rho_G
=
\operatorname{Tr}_F\left[\ketbra{\Psi}_{FG}\right]
\end{equation}
contains off-diagonal coherence in the computational basis, and the success probability depends on these coherences.

\subsection{Exact success probability in the general case}

Let $S_{\Lambda}\subseteq \mathbb{F}_2^{N_g}$ denote the affine support
span of the populated garbage support $\Lambda$ with
$\dim S_{\Lambda}=r_{\Lambda}$, let
$Q=\mathbb{F}_2^{N_g}/S_{\Lambda}^\perp$ be the effective syndrome
space, identified with $\mathbb{F}_2^{r_\Lambda}$ through the syndrome
map $q(k)$ of \cref{eqn:syndrome_vec}, and let
$\mathcal{A}\subseteq Q$ be an arbitrary accepted syndrome set.
For any $a\in Q$, choose an arbitrary representative
$\kappa_a\in q^{-1}(a)$. For $s\in S_{\Lambda}$, define the
Walsh--Hadamard coefficients of the indicator function of the accepted
syndrome set $\mathcal{A}$ by
\begin{equation}
\widehat{\mathbf{1}}_{\mathcal{A}}(s)
:=
\sum_{a\in\mathcal{A}}
(-1)^{\kappa_a\cdot s}.
\end{equation}
This definition is independent of the chosen representatives, because
changing $\kappa_a$ by an element of $S_{\Lambda}^\perp$ does not change
its inner product with any $s\in S_{\Lambda}$.

Note that the success probability is the probability that the $X$-basis
measurement outcome lies in
$q^{-1}(\mathcal{A})
\equiv\bigcup_{a\in\mathcal{A}}q^{-1}(a)$.
Therefore,
\begin{equation}
P_{\mathrm{succ}}(\mathcal{A})
=
\mathrm{Tr}\!\left(\Pi_{\mathcal{A}}\rho_G\right),
\qquad
\Pi_{\mathcal{A}}
:=
\sum_{k\in q^{-1}(\mathcal{A})}
\ketbra{k_X}{k_X}.
\end{equation}
For a fixed effective syndrome class $a\in Q$, the projector onto the
corresponding coset of $X$-basis outcomes is
\begin{equation}
\Pi_a
=
2^{-r_{\Lambda}}
\sum_{s\in S_{\Lambda}}
(-1)^{\kappa_a\cdot s}
X^s,
\label{eq:projector_general_onto_a}
\end{equation}
where $X^s:=\bigotimes_{i=1}^{N_g}X^{s_i}$.
Indeed, $\ket{k_X}$ is an eigenstate of $X^s$ with eigenvalue
$(-1)^{k\cdot s}$, so \cref{eq:projector_general_onto_a} projects
precisely onto those outcomes $k$ satisfying $q(k)=a$.
Summing \cref{eq:projector_general_onto_a} over $a\in\mathcal{A}$ gives
\begin{equation}
\Pi_{\mathcal{A}}
=
\sum_{a\in\mathcal{A}}\Pi_a
=
2^{-r_{\Lambda}}
\sum_{s\in S_{\Lambda}}
\widehat{\mathbf{1}}_{\mathcal{A}}(s)X^s.
\end{equation}
Taking the trace against $\rho_G$ therefore gives the success probability
associated with an accepted syndrome set $\mathcal{A}\subseteq Q$:
\begin{equation}
P_{\mathrm{succ}}(\mathcal{A})
=
2^{-r_{\Lambda}}
\sum_{s\in S_{\Lambda}}
\widehat{\mathbf{1}}_{\mathcal{A}}(s)
\mathrm{Tr}\!\left(X^s\rho_G\right).
\label{eq:succ_prob_general}
\end{equation}
Equivalently,
\begin{equation}
P_{\mathrm{succ}}(\mathcal{A})
=
\sum_{k\in q^{-1}(\mathcal{A})}
\bra{k_X}\rho_G\ket{k_X}.
\label{eq:succ_prob_general_2}
\end{equation}
\cref{eq:succ_prob_general_2} follows directly from the definition of
$\Pi_{\mathcal{A}}$ in the $X$ basis.

In the branch-orthogonal case, $\rho_G$ is diagonal in the computational
basis, so $\mathrm{Tr}(X^s\rho_G)=0$ for every $s\neq0$, while
$\widehat{\mathbf{1}}_{\mathcal{A}}(0)=|\mathcal{A}|$. Hence,
\cref{eq:succ_prob_general} reduces to
$P_{\mathrm{succ}}(\mathcal{A})=|\mathcal{A}|/2^{r_{\Lambda}}$,
recovering \cref{eq:generalised-succ-prob}.

\subsection{Robustness to residual coherence}

\cref{eq:succ_prob_general} shows that only coherences aligned with the
affine support span $S_{\Lambda}$ can modify the success probability.
Moreover, for a general accepted syndrome set $\mathcal{A}$, these
coherences are weighted by its Walsh--Hadamard coefficients.

Define the acceptance-weighted coherence
\begin{equation}
C_{S,\mathcal{A}}(\rho_G)
:=
\sum_{s\in S_{\Lambda}\setminus\{0\}}
\left|
\widehat{\mathbf{1}}_{\mathcal{A}}(s)
\right|
\left|
\mathrm{Tr}(X^s\rho_G)
\right|.
\end{equation}
Then the deviation from the success probability of the branch-orthogonal
case (\cref{eq:generalised-succ-prob}) satisfies
\begin{equation}
\left|
P_{\mathrm{succ}}(\mathcal{A})
-
\frac{|\mathcal{A}|}{2^{r_{\Lambda}}}
\right|
\le
2^{-r_{\Lambda}}
C_{S,\mathcal{A}}(\rho_G).
\label{eq:deviation_nonorthogonal}
\end{equation}
Indeed, the $s=0$ term in \cref{eq:succ_prob_general} equals
$|\mathcal{A}|/2^{r_{\Lambda}}$, and the bound follows by applying the
triangle inequality to the remaining terms.

This bound can also be expressed in terms of the trace distance from the
completely dephased garbage state. Let $\Delta$ denote complete dephasing
in the computational basis of the garbage register:
\begin{equation}
\Delta(\rho_G)
=
\sum_y
\ket{y}\!\bra{y}\rho_G\ket{y}\!\bra{y}.
\end{equation}
Since $\mathrm{Tr}(X^s\Delta(\rho_G))=0$ for all nonzero $s$, we have
$\mathrm{Tr}(X^s\rho_G)
=\mathrm{Tr}[X^s(\rho_G-\Delta(\rho_G))]$ for $s\neq0$.
H\"older's inequality and $\|X^s\|_\infty=1$ then give
\begin{equation}
\left|
\mathrm{Tr}(X^s\rho_G)
\right|
\le
\|\rho_G-\Delta(\rho_G)\|_1.
\end{equation}
Substituting this into \cref{eq:deviation_nonorthogonal} yields
\begin{equation}
\left|
P_{\mathrm{succ}}(\mathcal{A})
-
\frac{|\mathcal{A}|}{2^{r_{\Lambda}}}
\right|
\le
\Gamma_{\mathcal{A}}
\|\rho_G-\Delta(\rho_G)\|_1,
\label{eq:deviation_norm_bound}
\end{equation}
where
\begin{equation}
\Gamma_{\mathcal{A}}
:=
2^{-r_{\Lambda}}
\sum_{s\in S_{\Lambda}\setminus\{0\}}
\left|
\widehat{\mathbf{1}}_{\mathcal{A}}(s)
\right|.
\end{equation}

For the pure post-selection case $\mathcal{A}=\{0\}$, every
Walsh--Hadamard coefficient equals one,
$\widehat{\mathbf{1}}_{\{0\}}(s)=1$ for all $s\in S_{\Lambda}$, so
\cref{eq:deviation_nonorthogonal,eq:deviation_norm_bound} reduce to
\begin{align}
\left|
P_{\mathrm{succ}}
-
2^{-r_{\Lambda}}
\right|
&\le
2^{-r_{\Lambda}}
\sum_{s\in S_{\Lambda}\setminus\{0\}}
\left|
\mathrm{Tr}(X^s\rho_G)
\right|
\nonumber\\
&\le
(1-2^{-r_{\Lambda}})
\|\rho_G-\Delta(\rho_G)\|_1 .
\end{align}

These bounds show that MBU is robust to residual coherence in the garbage
register. Only coherence components aligned with the affine support span
contribute to the success probability, and for a general accepted set
even these contributions are further filtered by the Walsh--Hadamard
coefficients $\widehat{\mathbf{1}}_{\mathcal{A}}(s)$ of $\mathcal{A}$;
all other residual coherence has no influence on the success probability.

\end{document}